\documentclass[10pt,journal,compsoc]{IEEEtran}
\ifCLASSOPTIONcompsoc
  \usepackage[nocompress]{cite}
\else
  \usepackage{cite}
\fi

\ifCLASSINFOpdf
 
\else
 
\fi
\usepackage{microtype}                 % use micro-typography (slightly more compact, better to read)
\PassOptionsToPackage{warn}{textcomp}  % to address font issues with \textrightarrow
\usepackage{textcomp}                  % use better special symbols
\usepackage{mathptmx}                  % use matching math font
\usepackage{times}                     % we use Times as the main font
\usepackage{cite}                      % needed to automatically sort the references
\usepackage{tabu}                      % only used for the table example
\usepackage{booktabs}  
\usepackage{multirow}
\usepackage{multicol}
\usepackage[table,xcdraw]{xcolor}
\usepackage[pdftex]{graphicx}
\usepackage[normalem]{ulem}
\useunder{\uline}{\ul}{}
\usepackage{flushend}

\begin{document}

\title{Finding Patterns in Visualized Data by Adding Redundant Visual Information}

\author{Salomon~Eisler
        and~Joachim~Meyer,~\IEEEmembership{Senior~Member,~IEEE}
         
\IEEEcompsocitemizethanks{\IEEEcompsocthanksitem S. Eisler is a Ph.D. student at the Department of Industrial Engineering at Tel Aviv University\protect\\
% note need leading \protect in front of \\ to get a newline within \thanks as
% \\ is fragile and will error, could use \hfil\break instead.
E-mail: salomone@mail.tau.ac.il
\IEEEcompsocthanksitem J. Meyer (M’01-SM’10) is Celia and Marcos Maus Professor for Data Sciences at the Department of Industrial Engineering, Tel Aviv University, Tel Aviv, Israel. E-mail:jmeyer@tauex.tau.ac.il}% <-this % stops an unwanted space
\thanks{Manuscript received April 19, 2022; revised August 26, 2022.}}

% The paper headers
\markboth{Journal of \LaTeX\ Class Files,~Vol.~??, No.~?, August~2022}%
{Eisler \MakeLowercase{\textit{et al.}}: Finding Patterns in Visualized Data by Adding Redundant Visual Information}

\IEEEtitleabstractindextext{%
\begin{abstract}
We present "PATRED", a technique that uses the addition of redundant information to facilitate the detection of specific, generally described patterns in line-charts during the visual exploration of the charts. We compared  different versions of this technique, that differed in the way redundancy was added, using nine distance metrics (such as Euclidean, Pearson, Mutual Information and Jaccard)  with judgments from data scientists which served as the "ground truth". Results were analyzed with correlations ($R^2$), F1 scores and Mutual Information with the average ranking by the data scientists. Some distance metrics consistently benefit from the addition of redundant information, while others are only enhanced for specific types of data perturbations. The results demonstrate the value of adding redundancy to improve the identification of patterns in  time-series data during visual exploration.
\end{abstract}

}

\maketitle

\IEEEdisplaynontitleabstractindextext

\IEEEpeerreviewmaketitle

\IEEEraisesectionheading{\section{Introduction}\label{sec:introduction}}
\IEEEPARstart{D}{ata}science practitioners routinely explore the models they develop ~\cite{wexler2019if}, using visualizations to understand and to improve them. Data practitioners, in general, often scrutinize key structures of time-series visualizations \cite{brockwell2002introduction} to search for specific patterns ~\cite{aigner2011visualization}, to discover motifs ~\cite{esling2012time}, to detect anomalies and outliers and to predict ~\cite{esling2012time}.  

\begin{figure}[ht]
		\centering
		\includegraphics[width = 5cm]{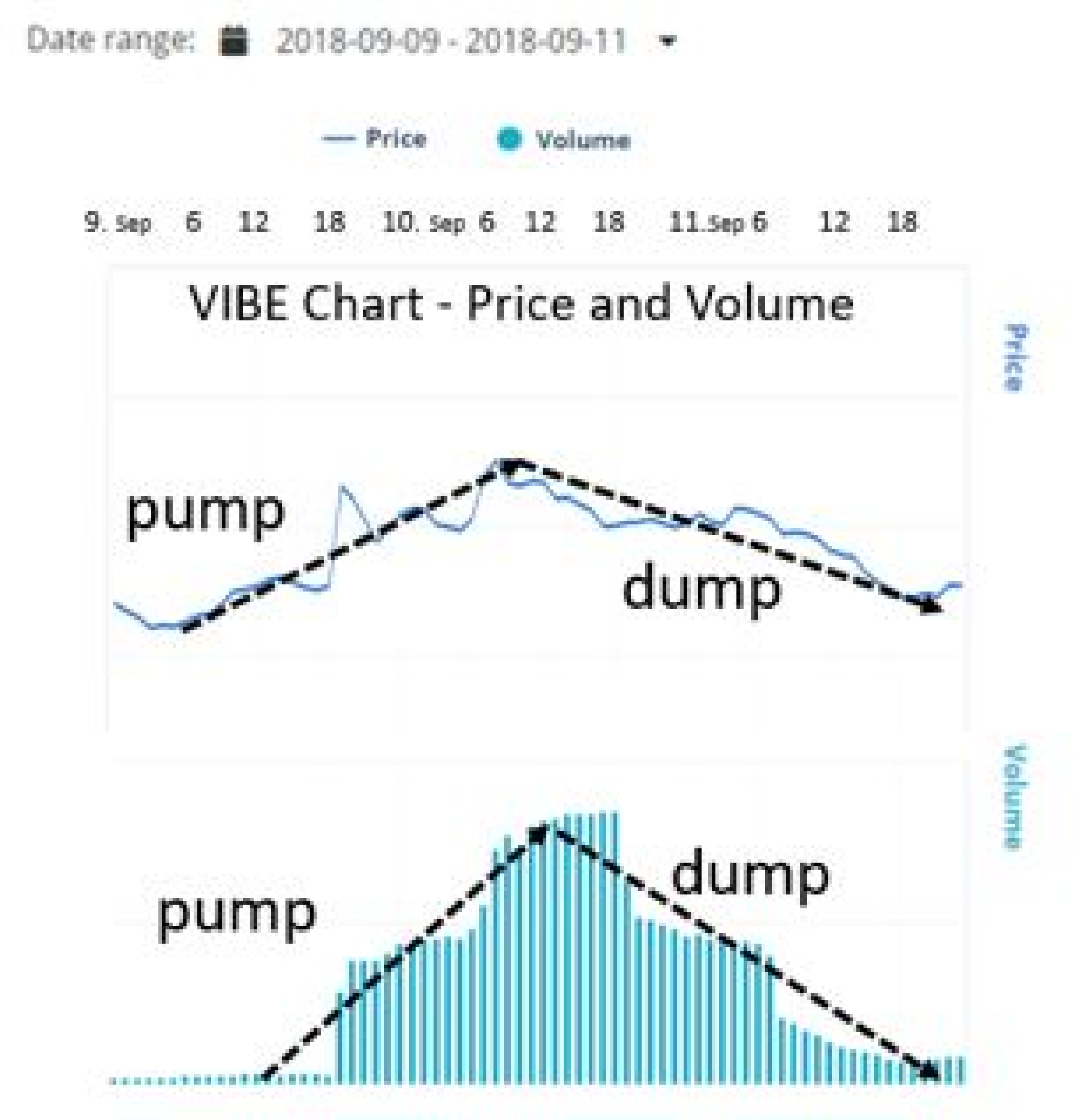}
		\caption{Vibe prices and volume from coincheckup.com}
		\label{vibe} 
		\end{figure}

The identification of patterns and motifs in time-series visualizations is complex, because events differ in properties, such as their evolution over time, the magnitude of changes, or the exact shapes of the patterns ~\cite{esling2012time}. However, despite the complexity, people still try to identify specific patterns in visualized data ~\cite{esling2012time}. For example, when inspecting visualizations of stock trading or other financial time-series ~\cite{wan2017formal}, analysts can point to events that seem to match a specific pattern, such as episodes of "pump and dump" in crypto-currency markets. These are market manipulation events that consist of artificially inflating the price of an owned security (in this case a crypto-currency) and then selling it at a much higher price ~\cite{la2020pump}.  Figure ~\ref{vibe} shows the hourly price for Vibe from September 9 to 11, 2018. As can be seen, orders are scarce during the hours before the pump and suddenly grow when the scheme begins. The sudden increase in volume is a pattern that resembles a classical pump and dump episode, including the fact that the  volume quickly returned to the initial level (the dump) ~\cite{la2020pump}.
In general, analysts may look for and try to identify various patterns, with the specifics depending on the data domains (e.g. cardiology, finance, intrusion detection, etc.). In particular, in the financial area, 53  different time-series patterns were reported in the literature ~\cite{wan2017formal}.

Large time series can contain hundreds of different patterns, so it may be necessary to support interactive visual analyses and exploration ~\cite{hao2012visual}. One of the main challenges in the Visual exploration of patterns in time-series is the efficient search for patterns that are similar (i.e. low distance) to a specific motif ~\cite{hao2012visual}. We explore the effects of adding redundancy to the displayed data on the ability to detect patterns in data series, given an user-created input that resembles the pattern. We also aim to determine which distance metrics which support similarity detection benefit more from redundancy.  

To support the visual exploration process, we developed "PATRED" (from "PATterns" and "REDundancy"), a technique that uses redundant visual information to improve the detection of specific patterns or motifs in time-series line-charts visualizations. We compared  different versions of this technique that differed in the way redundancy was added. We also introduced perturbations in the exemplar data sets and created somewhat different data. We used nine methods (such as Pearson correlations, Mutual Information and Jaccard) to compute the relation between the exemplar data and the displayed data for the different redundancy - perturbation combinations. Judgments from data scientists served as the "ground truth" to which we compared the output from the different methods. The results showed that distance metrics which are based on data groupings (e.g. Mutual Information, Jaccard, Dice, Cosine) consistently had higher $R^2$ and F1 score values with PATRED than without PATRED when predicting data scientists' judgments. Metrics which are based on distances between individual data points (e.g. Pearson, Euclidean, Manhattan) benefited much less from the added redundancy.

\begin{figure}[ht]
		\centering
		\includegraphics[width = 8.5 cm]{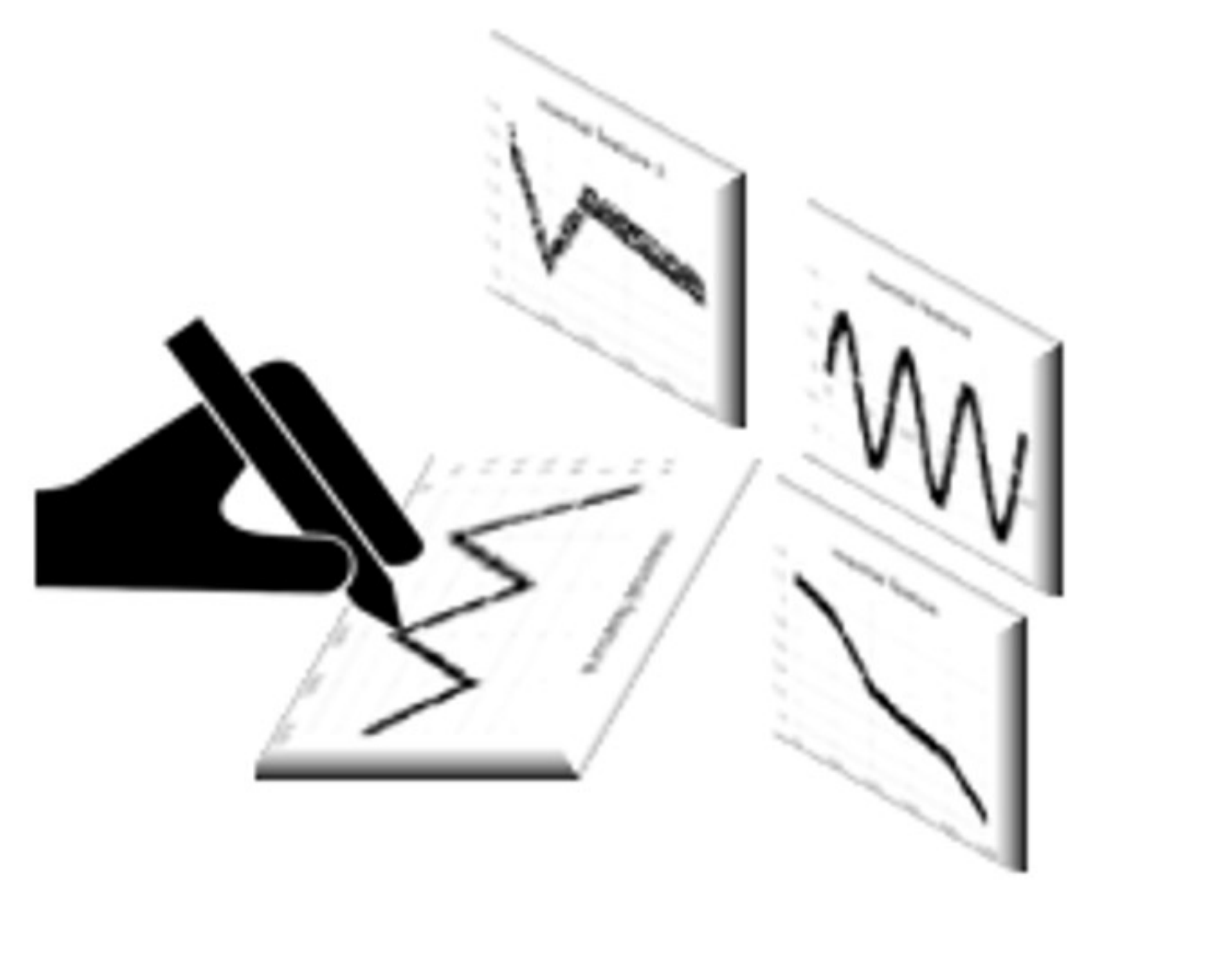}
		\caption{A data scientist digitizes the sought patterns}
		\label{digitize1} 
		\end{figure}

To use this method, and to start the exploration process, practitioners first need to provide PATRED with rough examples of the pattern(s) they are interested in (see Figure ~\ref{digitize1}), for instance by sketching the pattern.  PATRED then adds redundancy to both the data from the pattern examples and the explored data. It then computes the distance metrics between the examples and the data set to identify the groups with the lowest distance metric. 

PATRED can be used in various types of exploratory data analysis  \cite{aigner2011visualization} such as 
i) pattern definition, pattern localization and pattern/motif discovery, ii) relation seeking and pattern comparisons, iii) identification of unusual patterns (anomaly  detection), and iv) support for clustering analyses. In this paper we will focus on i. Tasks of types ii,iii and iv will be addressed in future work. 

In section  2 below,  we present a short overview of relevant work in this area and provide some background on Information Theory. In section 3 we review different methods for measuring similarities/distances between variables which can be used to identify patterns. In section 4 we describe our method, using redundant visual information to search for patterns.  In section 5 we describe a study we conducted with data scientists who ranked similarities between charts.  In section 6 we compare different approaches, applying the method to different cases and show why we recommend to include it in the pattern recognition tool box. In section 7 we summarize the results and limitations. In section 8 we provide some conclusions and discuss future work and practical applications of our approach in data science and data exploration.

\section{Background and Relevant Work}\label{sec:background}

We first present a short overview of relevant work and provide some background on  patterns, Information Theory and other related concepts that will be used in the following sections.

\subsection{Patterns}

Pattern identification in line charts involves the simultaneous consideration of multiple data-points and sometimes the whole chart. Data practitioners interpret the data behavior and assign a pattern to the visualized data, based on their understanding of the visualization  ~\cite{andrienko2006exploratory}. Different people may see different patterns in the same set of data.

The purpose for which the data is analyzed and the data practitioners' goals determine the patterns they look for in a given situation.  Patterns will frequently be described in a general, inexact manner, rather than with precise specifications ~\cite{andrienko2006exploratory}.

\subsection{Visualization and Information Theory}
A communication or transmission channel is the physical medium through which information is transmitted, such as telephone lines, or atmosphere in the case of wireless communication \cite{chaudhary2011error}. Engineers and researchers aim to maximise the proportion of the entropy of the output that reflects information from the input \cite{stone2015information}. This is the Mutual Information between the input and the output of the channel. 

\begin{figure}[ht]
		\centering
		\includegraphics[width = 8.5 cm]{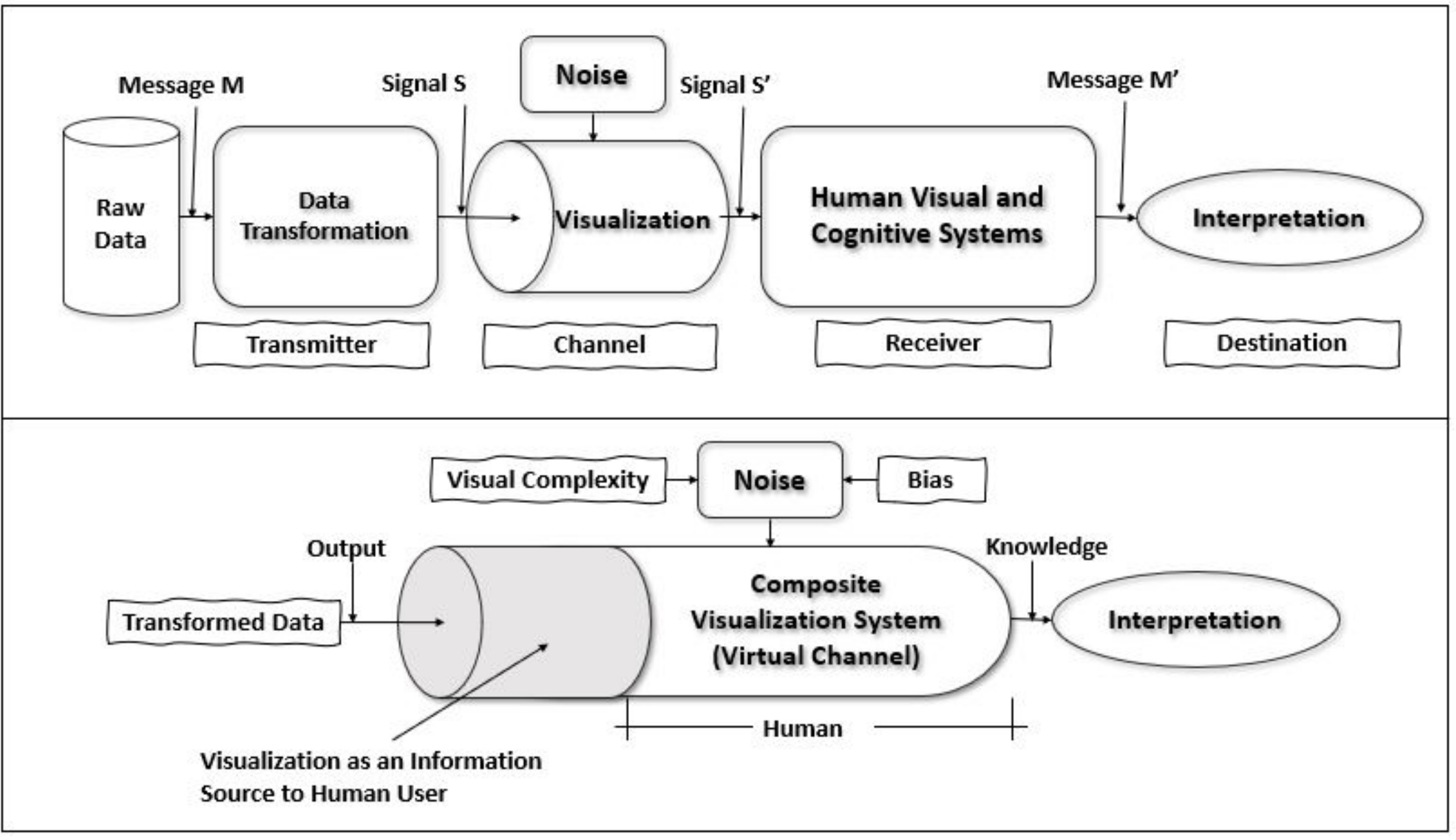}
		\caption{Visualizations as communication channels}
		\label{visualization_as_channels} 
		\end{figure}

It is possible to use Information Theory  \cite{shannon2001mathematical} to describe the visualization process by referring to the data transformation, the visualization and the human as components of a communication channel. In the upper part of Figure ~\ref{visualization_as_channels} we show a diagram of the basic communication system considered by Shannon \cite{shannon2001mathematical}, referring to the Data Transformation as the Transmitter, the Visualization as the Channel and the Human Visual and Cognitive Systems as the Receiver.  Similarly, the process of interacting with information visualizations can be viewed as a virtual communication channel \cite{Chen2010}, composed of a series of channel segments, from the data source to the human viewer.
The bottom part of Figure ~\ref{visualization_as_channels} depicts this virtual information channel, where its last segment is the human (before the destination), and the visualization is the source. 
The data transformation segment can be just a simple aggregation process or a more complex process which includes algorithms such as t-SNE (t-distributed stochastic neighbor embedding) or PCA (Principal Component Analysis).

The application of Information Theory to visualizations and related topics is not new \cite{Chen2010,conant1976laws,Card1999,purchase2008theoretical,chen2018cost}. For instance, \cite{Biswas2013} introduced a framework for exploring multivariate data sets to guide users through the multivariate data exploration process, using the concept of Mutual Information. Shannon’s entropy has also been mentioned as one of the best ways to evaluate human interaction with visualizations \cite{mcnamara2016entropy}.  It is possible to explain a number of phenomena in visualizations with Information Theory \cite{Chen2010}. 

Other studies examined different methods to identify similarity between images and charts, using correlations and mutual entropy \cite{reshef2011detecting,bermudez2018analysing}.

In the next section we will discuss several distance  metrics to enable detection of relations and similarity between variables, using Mutual Information, Pearson correlations and other metrics. 

\section{Distance metrics for similarity detection between variables }
The evaluation of the relation between two or more random variables has been studied for more than a century by researchers such as Galton, Pearson and others \cite{smith2015mutual}, leading to the development of numerous metrics \cite{yu2006new}. For instance, the R package philentropy alone can compute 46 different distance and similary metrics (i.e. distance is the complement of similarity) and 50 are mentioned in ~\cite{cha2007comprehensive}, grouped in different distance families. Known methods are Manhattan and Euclidean distances ($L_{p}$ Minkowski family),  Cosine (Inner Product family), Jaccard and Dice similarities (Intersection Family), Pearson correlation (Squared   $L_2$ family) and Mutual Entropy (Shanon Entropy family).  Additional details regarding the Manhattan, Euclidean, Cosine, Jaccard, Dice and Pearson distances can be found in the Appendix.

Two important similarity metrics are based on information theory and belong to the Shannon’s entropy family. They are the Normalized Mutual Information (NMI) and the relative entropy or Kullback-Leibler divergence (KL) \cite{cover2012elements}.  The term entropy is a measure of information \cite{shannon2001mathematical}. For any probability distribution  $p(x)$ \cite{cover2012elements}, the entropy is defined as Equation \ref{eqentropy} below:

\begin{equation}\label{eqentropy}
\ Entropy\; H(X) = -\sum_{i=1}^{N}p(x_{i})\log p(x_{i})\    
\end{equation}

\begin{figure}[ht]
	\centering
	\includegraphics[width = 8.5cm]{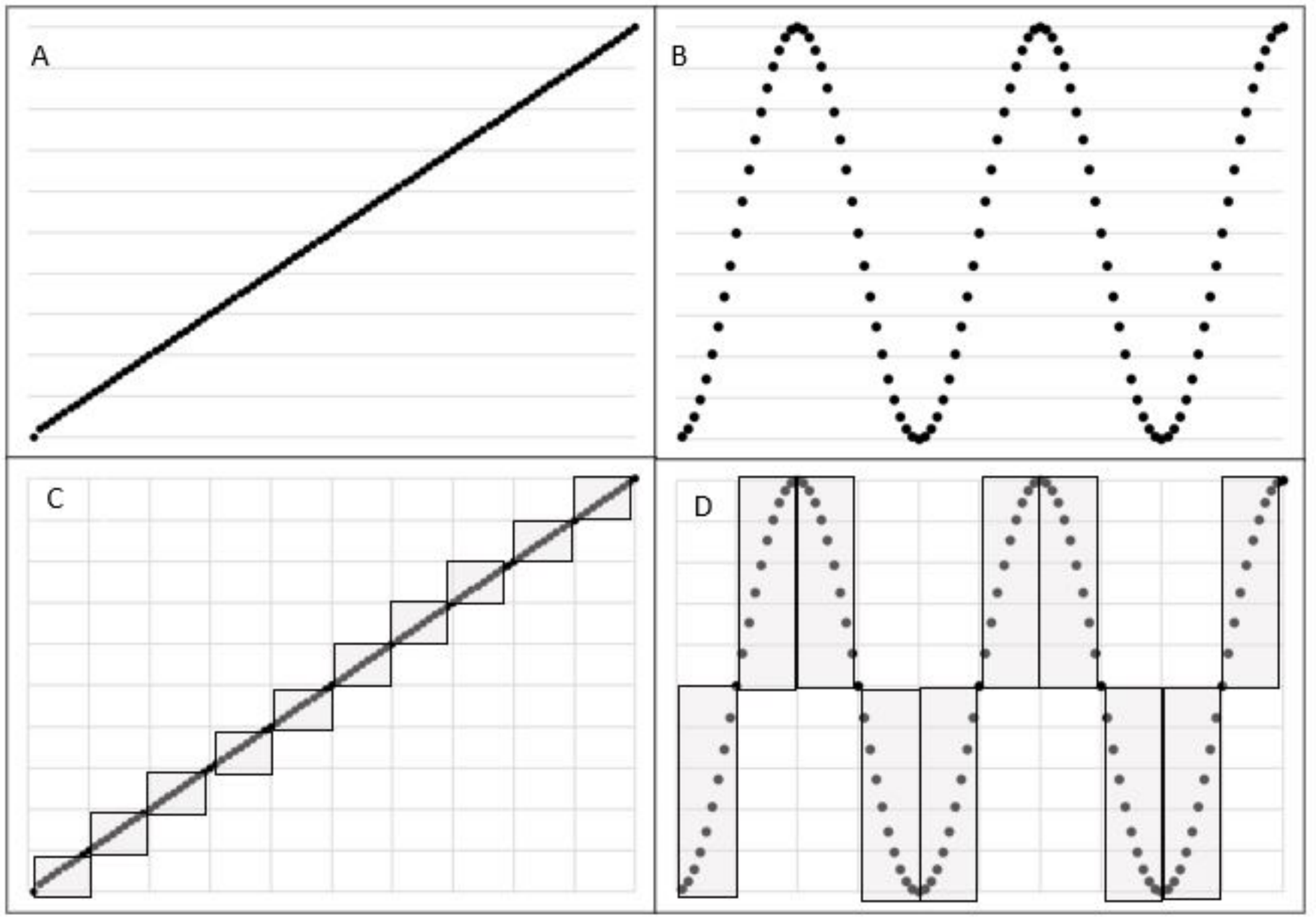}
	\caption{Illustration of multiple bins used to calculate entropy. If data is divided only along the y axis, as in A and B, the entropy for A and B is the same. In C and D, the data is divided along both the x and y axes, and it is possible to differentiate between different structures in the data. This type of bins configuration helps when comparing images with different methods.}
	\label{mays_entropy} 
	\end{figure}

Where i is the bin number, $N$ is the number of populated bins, and $p(x_i)$  is the proportion of data falling into the i-th. populated bin, subject to the condition $\Sigma{p(x_i)} = 1$. The number $N$ is determined by the size of the the histogram bins used to compute the Probability Density Function (PDF) from the observed data \cite{mays2002information}, see Figure  ~\ref{mays_entropy} for a simple example of histogram bins. The log is to base 2 and entropy is expressed in bits. If the PDF is peaked, the data is predictable and the information entropy is low. If the PDF is uniform, the data is unpredictable, and the information entropy is high. 

Mutual Information, an expansion of the entropy concept, is the amount of information one random variable conveys about another random variable \cite{cover2012elements}. It measures the generic association between two variables, such as the input and output of a communication channel \cite{stone2015information}. Equation \ref{eqmutual} shows the relation of the Mutual Information between two variables $I(X,Y)$, the entropy of each variable $H(X)$ and $H(Y)$, and their joint entropy $H(X,Y)$ \cite{cover2012elements}.

\begin{equation}\label{eqmutual}
\ {I\left(X,Y\right)=H\left(X\right)+H\left(Y\right)-H\left(X,Y\right)} \;  
\end{equation}

Since the Mutual Information (or Mutual Entropy) can be any positive number, it is useful to normalize it to obtain a value in the interval $[0,1]$. Equation \ref{eqn_mutual} provides a definition of $NMI(X,Y)$.

\begin{equation}\label{eqn_mutual}
\ Normalized(MI)\;NMI(X,Y) =  \frac{I(X,Y)}{\sqrt{H(X)H(Y)}}
\end{equation}

When $NMI(X,Y)$ increases from 0 to 1, the similarity or relation between between $X$ and $Y$ moves from fully dissimilar/independent to fully similar/dependent. This is analogous to the behavior of the Pearson correlation coefficient, where its value is 0 for fully dissimilar/independent variables and 1 (or -1) for fully dependent variables. Mutual Information \cite{cover2012elements} is less intuitive than the Pearson correlation, but it is more robust to effects of outliers ~\cite{correa2013mutual} and can help to detect non-linear associations.

Other methods, such as such as Taylor Diagrams (that evaluate the similarity or lack of similarity between time series, using Pearson Correlations and Standard Deviations) \cite{taylor2001summarizing} and Mutual Information Diagrams (MID) \cite{correa2013mutual} combine two or more metrics to detect similarity.
 
The Variation of Information (VI) between two variables X and Y is defined in Equation \ref{VIequa}.

\begin{equation}\label{VIequa}
\ {VI\left(X,Y\right)=H\left(X\right)+H\left(Y\right)-2I\left(X,Y\right)}\;  
\end{equation}

\begin{figure}[ht]
\centering
\includegraphics[width = 8.5cm]{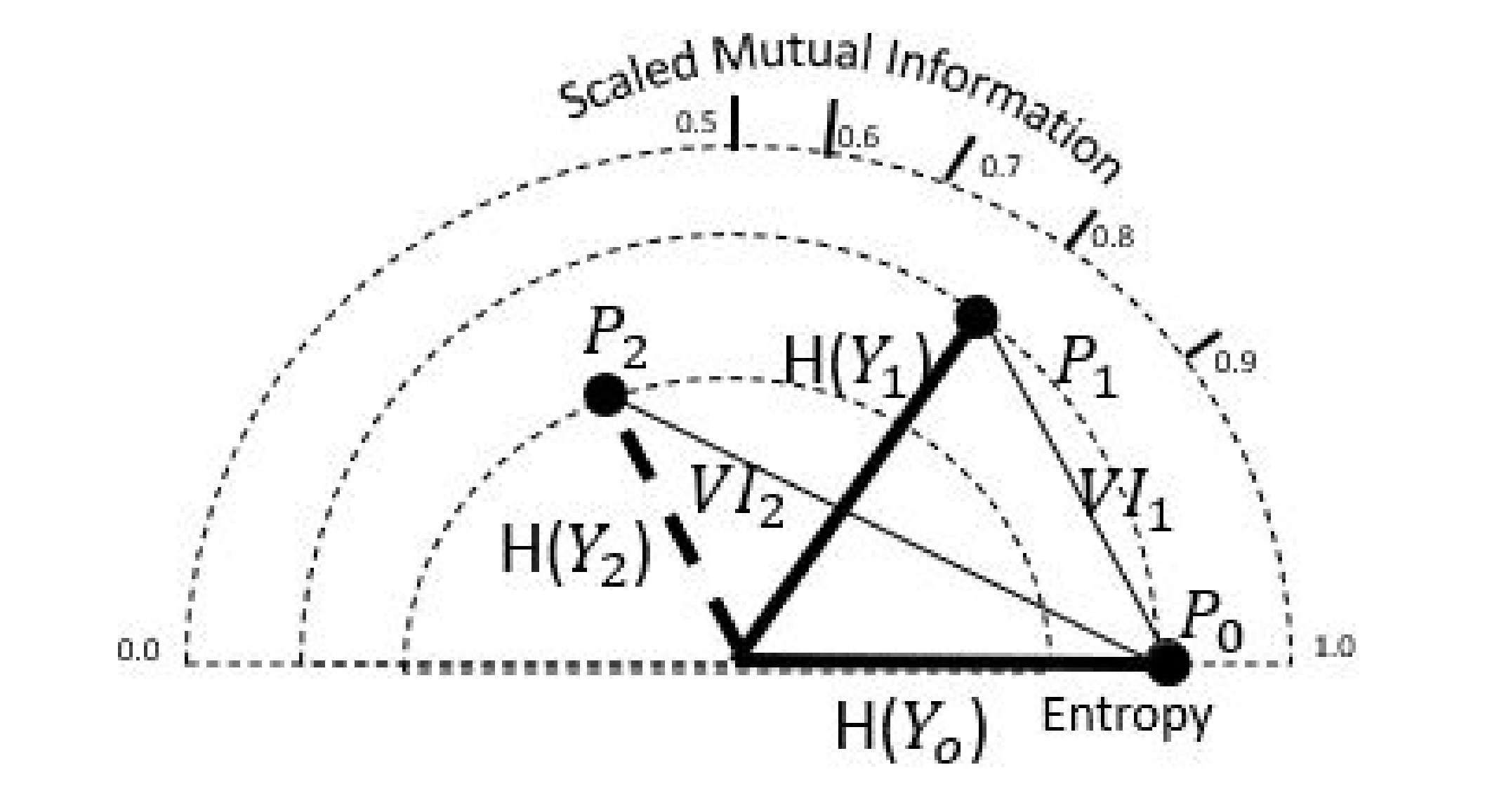}
\caption{Mutual Information Diagram \cite{correa2013mutual}  }
\label{MI_diag}
\end{figure}

It was used in \cite{correa2013mutual} to create Mutual Information Diagrams (MID) (see Figure ~\ref{MI_diag}), a visual method using two-dimensional plots that succinctly show the relations between two or more random variables. Variation of Information (VI) uses entropies and mutual entropies (Equation \ref{VIequa}) instead of standard deviations and correlations. VI is used to derive triangle inequalities to generate a polar coordinates plot, using H(Y), H(X) and VI, which can be used to evaluate the similarity between variables.  The MID of a data set with itself is a point $P_o$ on the horizontal axis, which can be used as a reference point. In Figure  \ref{MI_diag} we expect that the distances $VI_1$ between $P_1$ and $P_o$ and $VI_2$ between $P_2$ and $P_o$ should be short when the data sets $Y_2$ and $Y_1$ are similar to $Y_o$. Distance metrics based on Mutual Information are best evaluated with MID distances \cite{correa2013mutual}.

The Kullback-Leibler divergence (KL) is calculated with Equation ~\ref{eqnkl}. Because it is the divergence of two probability distributions X and Y, it is also named Cross Entropy of X and Y ~\cite{cover2012elements}. Since the KL divergence is not symmetric, the Jensen-Shannon divergence (JSD) is often used instead. JSD is a symmetrized and smoothed version of the KL divergence. It is calculated with Equation \ref{eqnjsd}, where $M = \frac{1}{2}(X + Y)$.  
\begin{equation}\label{eqnkl}
\ KL(divergence)\;D(X||Y) = H\left(X,Y\right) - H\left(X\right)
\end{equation}
\begin{equation}\label{eqnjsd}
\ JSD(X||Y) =  \frac{1}{2}[D(X||M) +D(Y||M)] \end{equation}

\section{Overall Approach}
Our approach has two main components: we add redundancy to maximise the similarity that should be detected and use this to compute distances between images, rather than between time-series.  

\subsection{Adding redundancy to maximise similarity}
\begin{figure}[ht]
		\centering
		\includegraphics[width = 8.5 cm]{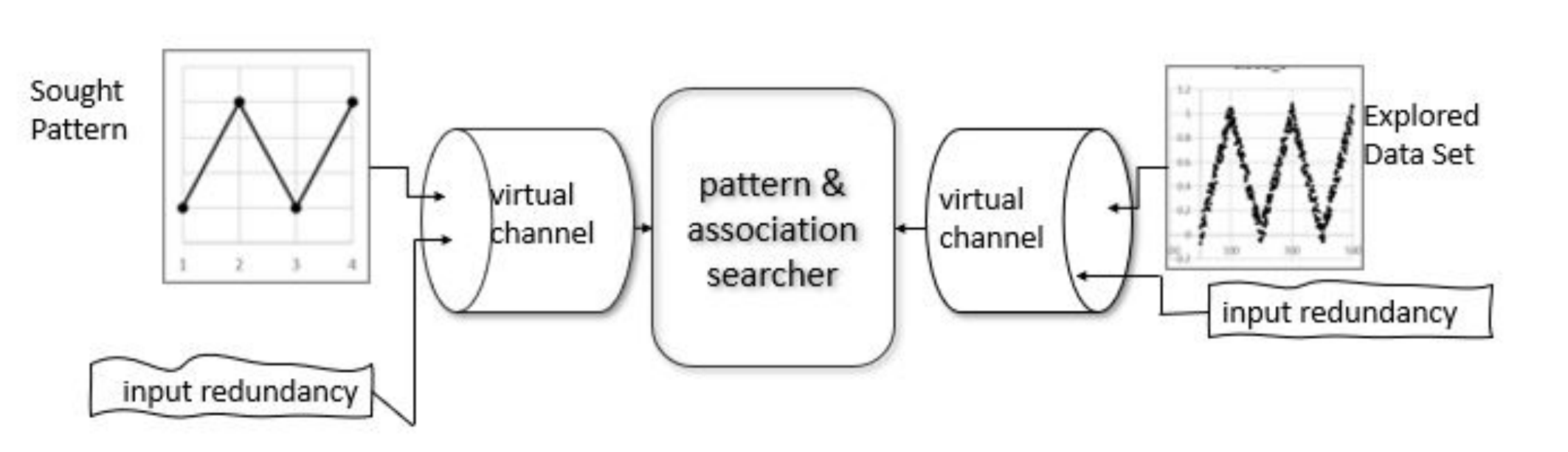}
		\caption{Adding redundancy to enhance similarity (Mutual Information). The pattern on the left is sought in the visualization on the right.}
		\label{visual_redundancy} 
		\end{figure}

We consider the process of identifying patterns within the data as two virtual communication channels that are connected through a decoder. The decoder tries to detect the similarity between the charts, as depicted in Figure ~\ref{visual_redundancy}, where both inputs are visualizations of data sets: on the left is the visualization of a pattern to be searched and on the right is the visualization of a larger data set to be explored, in which we expect to find the pattern. In Information Theoretic terms, for the decoder to detect similarity between the charts, it needs to detect high Mutual Information between them. 

Similar to what happens with physical communication channels, where undesirable disturbances (noise) can occur across the communication channel \cite{chaudhary2011error}, visualizations can contain noise, caused by bias or by the complexity of the visualization. Noise can cause errors in the decoding of the "messages". In our case, this means identifying incorrect patterns or failing to identify correct patterns.

As mentioned above, engineers and researchers aim to maximise the proportion of the entropy of the output that reflects information from the input \cite{stone2015information}. 

In our context, this means adjusting the data to maximise the similarity (Mutual Information), or minimize the distance, between the sought pattern and the inspected data set (see Figure ~\ref{visual_redundancy}).

\begin{figure}[ht]
		\centering
		\includegraphics[width = 8.5cm]{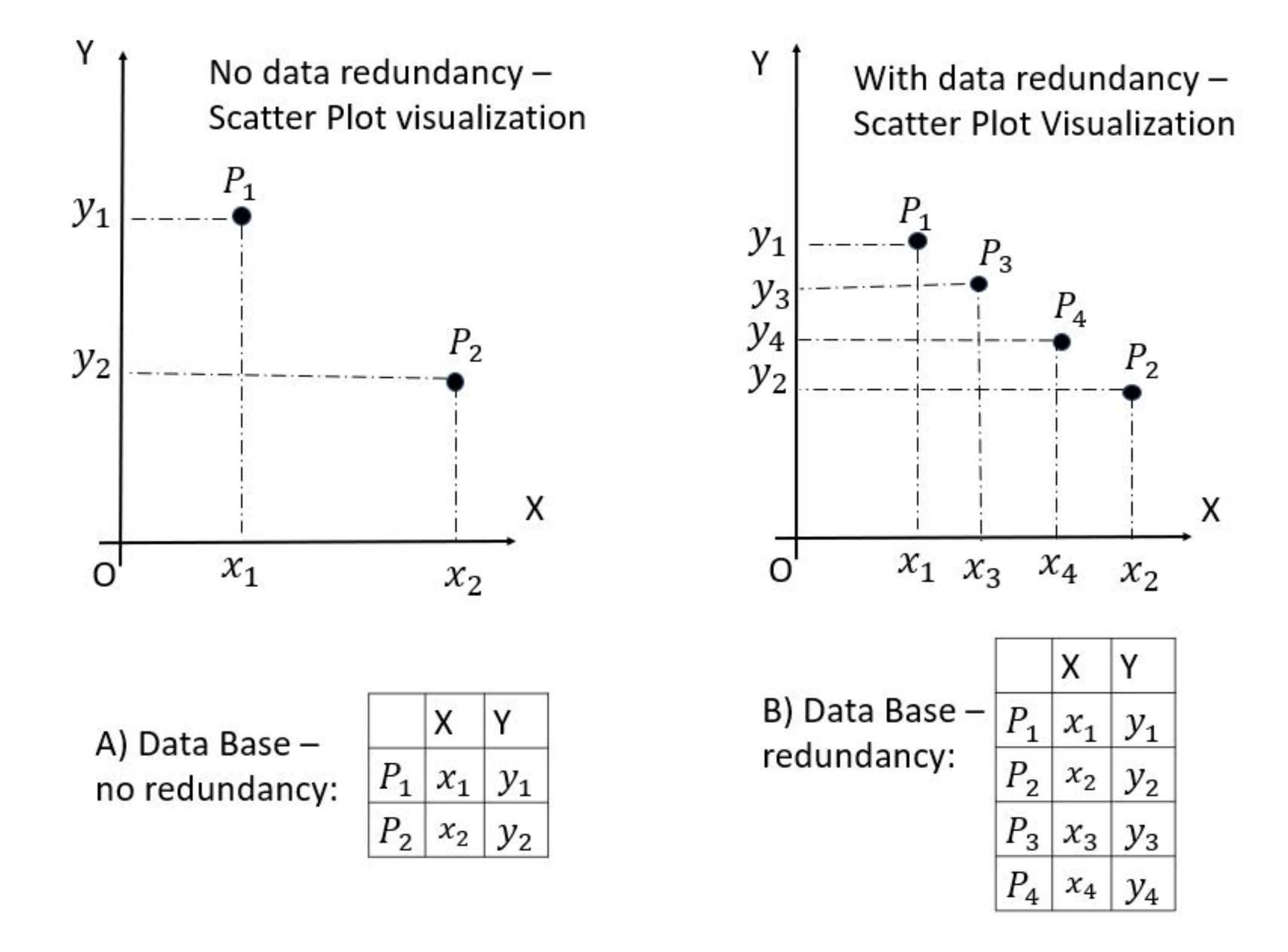}
		\caption{$P\_3$ and $P\_4$ are visual data redundancy they are on the line connecting $P\_1$ and $P\_2$}
		\label{simple_redundancy} 
		\end{figure}

A widely used method to make encoded messages more immune to the effects of noise is to add some redundancy to the signals before or while they are transmitted \cite{stone2015information}. The simplest way to add redundancy is to repeat the transmission several times, a practice called a repetition code  \cite{chaudhary2011error}. As with encoded messages, if redundancy is added without adding too much noise, the performance of the pattern detection can be improved. As with images and language, which are highly redundant, it is possible to add data points that are implicit in the existing data, thereby making the data more robust with respect to the effects of noise. Figure  ~\ref{simple_redundancy} depicts a very simple example of such a redundancy, where data points $P\_3$ and $P\_4$ are co-linear with $P\_1$ and $P\_2$. When a line is drawn between  $P\_1$ and $P\_2$,  $P\_3$ and $P\_4$ are invisible to the viewer, but for calculating the visual information, the addition of data points between $P\_1$ and $P\_2$ is helpful. 
%If too much redundancy is added, both sides of Figure  ~\ref{simple_redundancy} could become very similar, and the pattern detection would be worthless.

This "added redundancy" technique can also be used in other domains. One example is \cite{nakano1972noise}, where Information Theory concepts and redundancy were used to deal with "semantic noise", which is not disturbing the data signal but rather the meaning or message, transmitted with the signals.

\subsection{Adding redundancy to improve the distance metrics}
The addition of data values that are implicit in the rest of the data weakens the effects of visual noise. This is the basis for our first principle for redundancy design: it should be simple to recover the original pattern from the added data, to assure that the added data values are implicit in the original data.

A second point to consider is that our goal is to maximise the similarity (i.e. minimize the distance) between Y and X by adding entropy \cite{stone2015information}. Given that no probability distribution can have a larger entropy than the uniform distribution, we should maximise the uniformity of the redundant data. Therefore our second principle for redundancy design is to  maximise the uniformity of the redundant data.

%A third point to consider is not to add too much redundancy because then all the patterns would be similar one to the other.

\begin{figure}[ht]
		\centering
		\includegraphics[width = 8.5 cm]{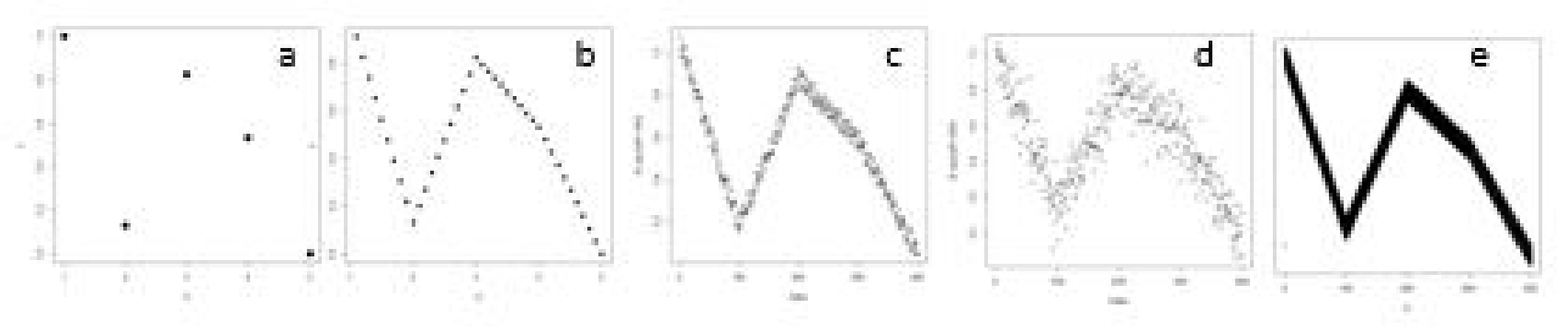}
		\caption{Examples of redundancy types for the same data set}
		\label{redun_clouds} 
		\end{figure}

Figure \ref{redun_clouds} depicts four examples of redundancy for a 4 period (5 dots) plot which is shown in (a).  Figure \ref{redun_clouds} (b) depicts a simple way to create redundancy by adding 10 points along each of the lines that connect two adjacent data points (similar to the example from Figure  ~\ref{simple_redundancy}). Clearly, both principles are applied in this example. First, it is easy to recover the original pattern after adding the redundancy. Second, the additional points are uniformly distributed between the original data points.  Figure \ref{redun_clouds} (c) shows a less obvious design for cloud redundancy. It starts with the redundancy from (b) with the addition of random uniform noise with $\eta$ in [0,0.1].  Figure \ref{redun_clouds} (d) is similar to (c), but it uses random Gaussian noise with $sd = 0.1$. Here the deviation from the design principles is larger, and so is the impact on the pattern identification.
In Figure \ref{redun_clouds} (e) we augmented the redundancy from (b) using both principles. First we added points between the original data points, and then we added 9 copies of the same line with vertical shifts of 0.01 between lines (for a total of 10 copies). In the rest of the paper we refer to this redundancy design as "Area Line". 

\subsection{Distance between images for time series}

\begin{figure}[ht]
		\centering
		\includegraphics[width = 8.5 cm]{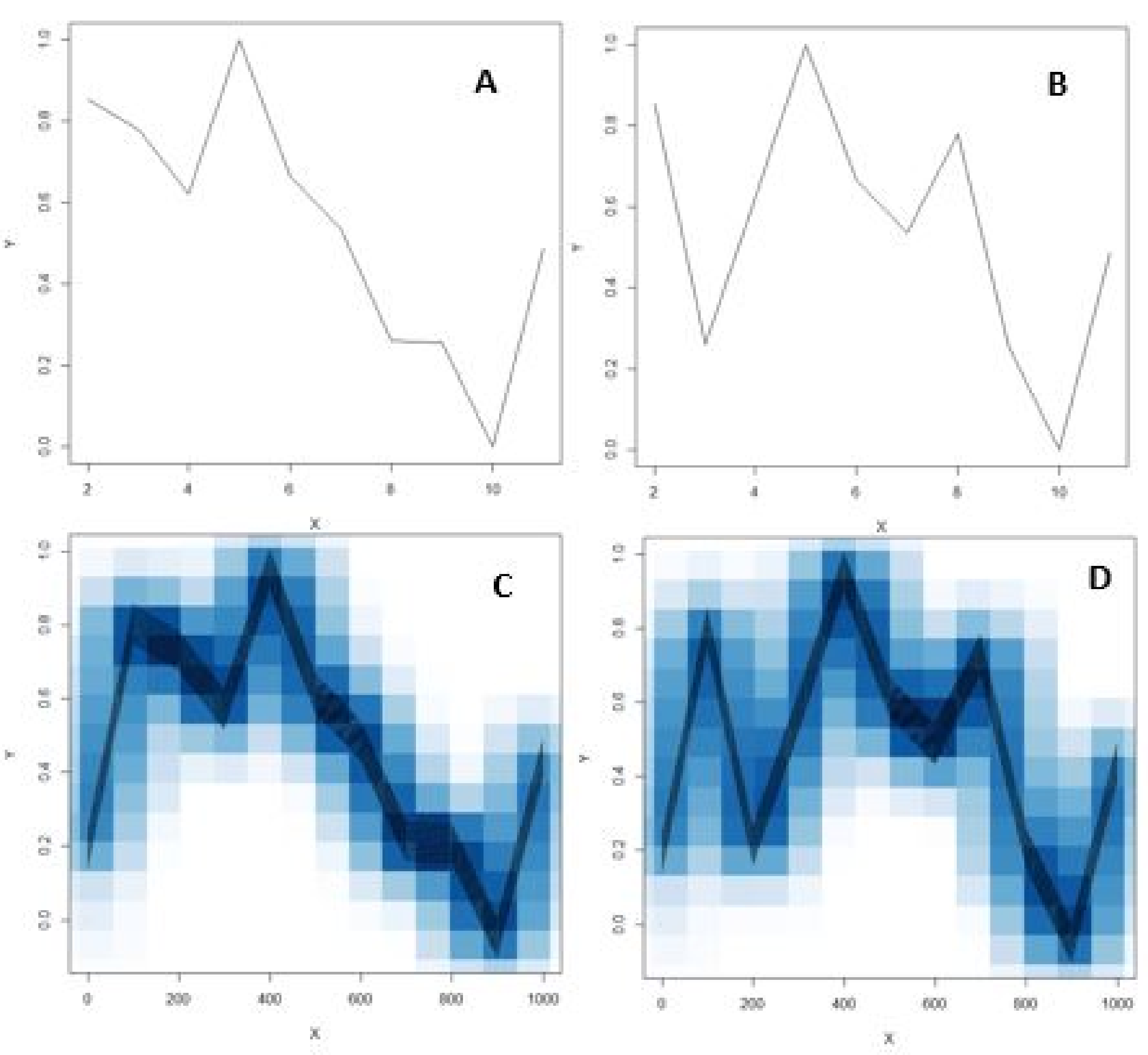}
		\caption{A, B are two standard line plots and C and D are two images in which the same line segments have width. The two axes were divided into 16 bins, so we have a 16 X 16 matrix of squares for the purpose of comparing the images.}
		\label{mi_images_ts} 
		\end{figure}

The usual way to compute the distance between two time-series is to calculate the distance between two 1D vectors of equal length, each containing a time-series that can be represented by one line plot  (Figure  ~\ref{mi_images_ts} A and B).  With it, the distance between X and Y can be computed, using the methods mentioned above.

Another option is to add width to each line plot, as seen in C and D in Figure  ~\ref{mi_images_ts}, and then to compute the distance between two images (C and D). The two axes were divided into 16 bins, so we have a 16 X 16 matrix of squares, which later can be used with the different distance metrics.
Similar approaches have been applied in other visualization domains. For instance, in medical imaging, Mutual Information is used to align ultrasound and x-ray images \cite{pluim2003mutual}.   Building on the experience from medical imaging, we can apply distance metrics between two images instead of between two time-series. Note that not all distance metrics may benefit from adding 2-dimensional redundancy, as will be discussed later.

\subsection{Redundancy configuration/methods for time-series line charts}

%Here we reiterate, more generally, what we showed above with the 4 examples of information redundancy in Figure \ref{redun_clouds}.

According to our first principle, to be able to fully recover the original data from the added data, the redundancy configuration must be consistent with the visual geometry of the data.  Based on this, two straightforward redundancy configurations would be: a) aligned equidistant points between each pair of adjacent data points and b) copies of a) with vertical shifts (generating broad lines between each original data point) (Figure ~\ref{redun_clouds} b and e).  There are other, more complex, alternatives for redundancy configuration, but these are outside the scope of this study.

\begin{figure}[ht]
		\centering
		\includegraphics[width = 8.5 cm]{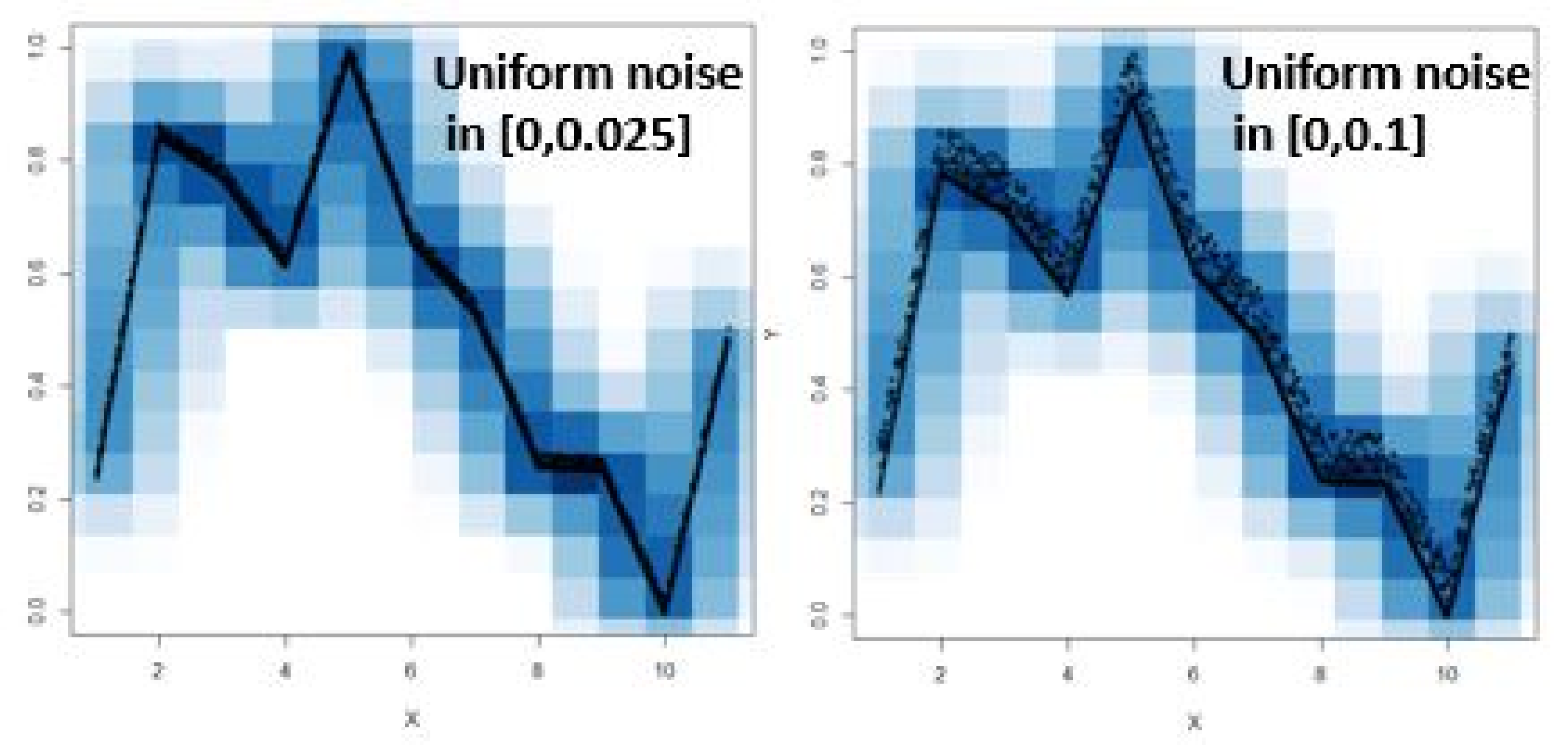}
		\caption{Cloud redundancy, with $\eta$ in [0,0.1] and $\eta$ in [0,0.025] with 100 redundant points, and square bins in blue, 8 per line segment.}
		\label{noise_red} 
		\end{figure}

In some cases it can help to add some uniform noise to the redundancy configurations, which makes the line chart look like a "cloud", aligned with the lines of the original chart. This type of "cloud" redundancy may resemble the internal representations of the target patterns (attentional templates) that are sought  ~\cite{kerzel2021attentional}, which are coarse and not very precise. We will refer to this kind of redundancy as cloud redundancy ( see Figure ~\ref{noise_red} for two examples). 

In our study, we used three combinations of information redundancy:  

\begin{enumerate}
    \item N redundancy without noise, (where N redundancy means N aligned equidistant points between each pair of adjacent data points) for N in $\{0,1,2,3,4,5,6,7,8,9,10, 15, 20, 25, 50,  75 \ and \ 100\}$.
    \item Area line  - 10 copies of N redundancy + vertical shifts of 0.01, for N in $\{0,1,2,3,4,5,6,7,8,9,10, 15, 20, 25, 50, 75 \ and \ 100\}$.
     \item Cloud -  (added random uniform noise $\eta$ in :[0,0.025],[0,0.1], [0,0.2]),  for N in $\{0,1,2,3,4,5,6,7,8,9,10, 15, 20, 25, 50,  75 \ and \ 100\}$.
\end{enumerate}

\section{Test of the effects of redundancy on results}

In order to compare the influence of the redundancy configurations on the results obtained with different distance metrics, we asked experienced Data Scientists to rank the visual similarities between time-series which had several controlled data perturbations. In the following subsections we describe the data perturbations scenarios and the survey with the Data Scientists.
In this section we used randomized selections of data from $nasdaq.com/market-activity/quotes/historical$ related to Apple daily traded volume between the dates: February 11, 2016 and September 12, 2017.

\subsection{Data perturbations scenarios }
\begin{figure}[!ht]
 \centering
 \includegraphics[width = 8cm ]{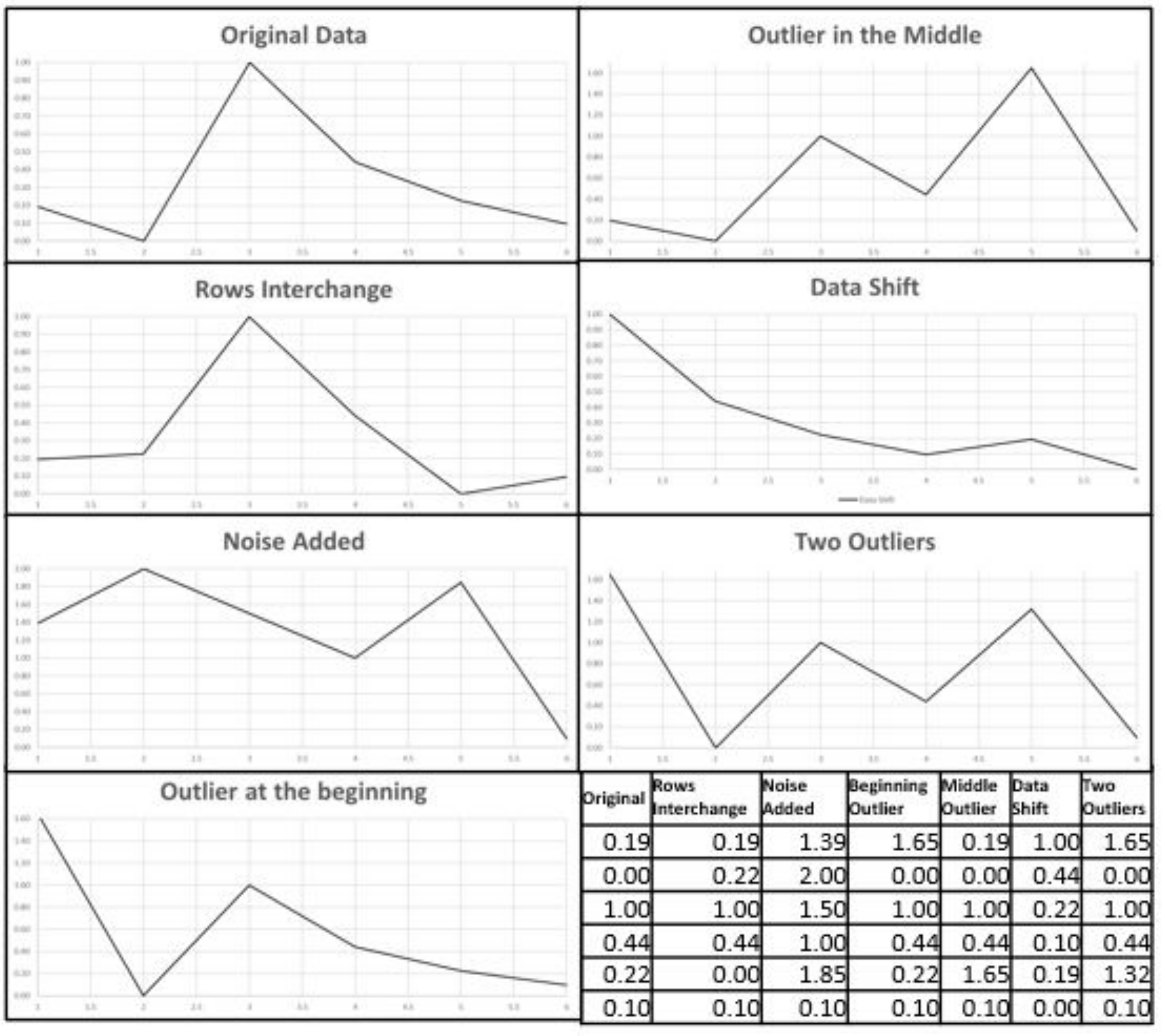}
 \caption{Six examples of data perturbations used for the similarity tests }
 \label{data_pertur}
 \end{figure}

We aimed to evaluate the effects from adding redundancy on the performance of pattern similarity detection tasks with different perturbations of the original data. The different perturbations of the original data were used to control the distance between two data sets. The starting point are two identical data sets, and we modify one of them by adding disturbances in the data. An example is shown in Figure ~\ref{data_pertur}, where the chart in the top left corner is the original data set and the other six charts depict the different data perturbations we used.

We applied three types of perturbations to the data: 1)Added noise, 2) Added outliers and 3) Data Shift. Noise is the most common case of data perturbation, and we introduced uniform random noise with values $[-1,1]$ to the original signal. Outliers are another common case, and different distance metrics may respond differently to them (e.g. Pearson is highly affected by Outliers). Data shifts are also frequent. In them, some sectors of the patterns remain the same and the rest are different. As benchmarks, we added three charts that were very different from the original data sets: Straight Line, Zigzag and Random Noise. 
Twelve data sets were created based on three randomly selected data sets of 6, 11, 16 and 21 data consecutive daily observations from NASDAQ. For each of the data sets, we introduced six types of data perturbations plus three charts with independent data sets (Straight Line, Zigzag and Random Noise). 
%The mutual entropy and the entropy of each original set (without perturbation) was computed for  each of the nine charts, applying the different methods mentioned above. 
The types of perturbations that were used are summarized in Table ~\ref{table:perturb_types}. These scenarios are not exhaustive, but they cover a large part of the real world cases.

\begin{table*}[ht]
\centering
\begin{tabular}{|l|l|}
\hline
Type                       & Data   Perturbation on the pattern                                                                                          \\ \hline
\multirow{4}{*}{Outliers}  & Interchange of 2   rows: interchanged 2 rows of the data without changing values.                                           \\ \cline{2-2} 
                           & Added 2 outliers: added 2 values that were 3 $sd$ above the mean of the   data set, without changing the order of the data. \\ \cline{2-2} 
                           & Added one outlier at the beginning of the data set (3 $sd$ above the   mean).                                               \\ \cline{2-2} 
                           & Added one outlier in the middle of the data set (3 $sd$ above the   mean).                                                  \\ \hline
Shift                      & Shifted the data set   2 periods right, and replaced the data for the two initial periods with the last 2 periods.        \\ \hline
Noise                      & Added uniform random   noise to the data.                                                                                   \\ \hline
\multirow{3}{*}{Benchmark} & Replaced the data   with a straight line.                                                                                   \\ \cline{2-2} 
                           & Replaced the data with a Zigzag line.                                                                                       \\ \cline{2-2} 
                           & Replaced the data with random uniform noise.                                                                                \\ \hline
\end{tabular}
	\caption{9 types of perturbation used to test the pattern rankings }
		\label{table:perturb_types} 
\end{table*}

\begin{figure*}[!ht]
 \centering
 \includegraphics[width = 18cm ]{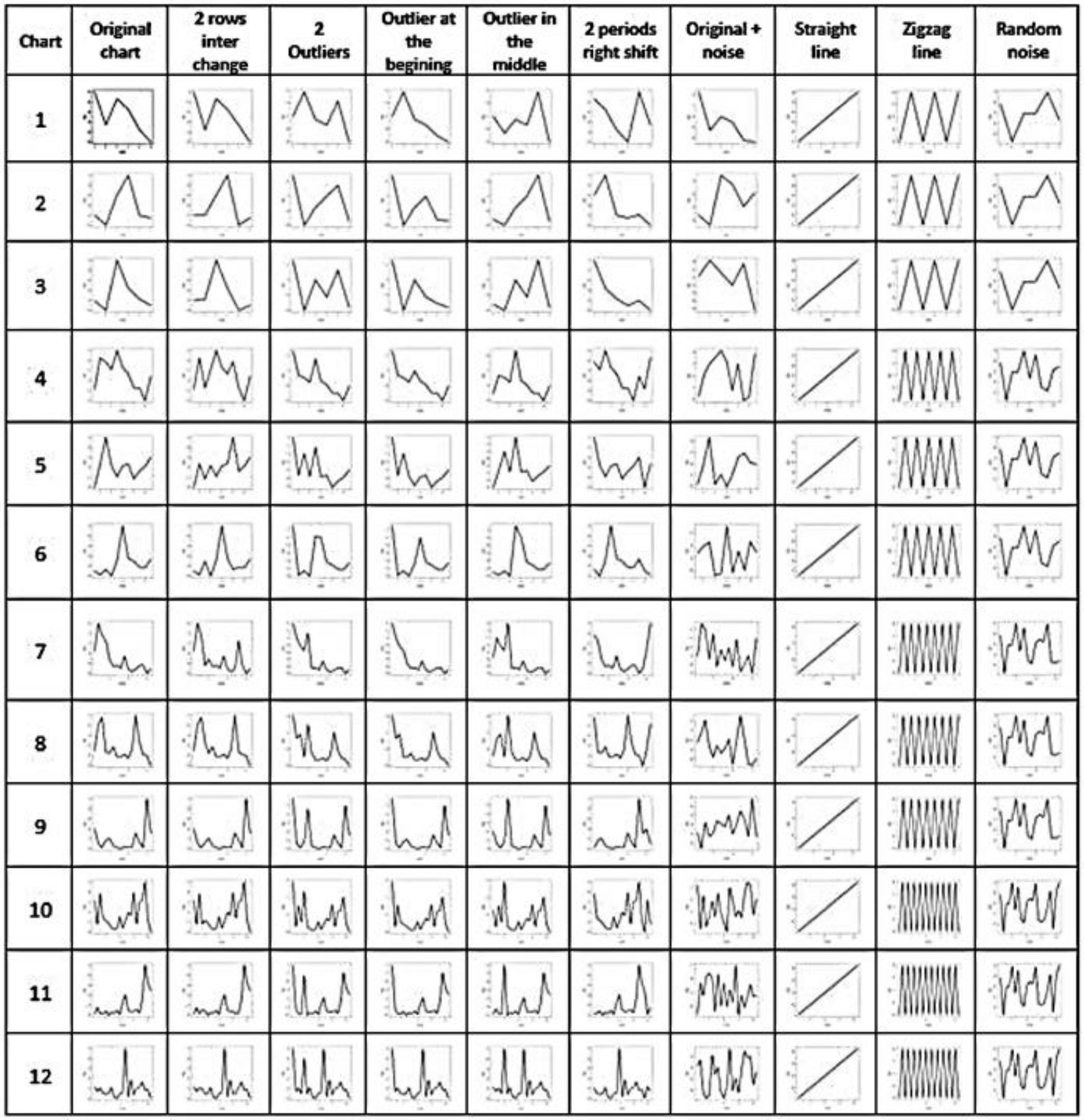}
 \caption{Twelve charts for the 12 data sets with 9 types of data perturbation}
 \label{data_sets}
 \end{figure*}

Figure \ref{data_sets} depicts the twelve charts for the 12 data sets with the different types of modifications. We aimed to determine how well each method assesses the similarities between the original charts and their modified versions.

\subsection{Data Science Practitioners as benchmark}

We conducted a survey with 13 experienced data scientists from the AI department in a gaming company. All of them had 3+ years of experience with M.Sc. or Ph.D. degrees. The survey was based on the data sets depicted in Figure \ref{data_sets} to compare the output from the different combinations of distance metrics and data redundancy to the data scientists' evaluations. We aimed to have a comparison benchmark for the "True" values of visual similarity judgements. 

The survey was anonymous and the data scientists did not have a time limit to provide their answers.  They only ranked the similarity (1 to 9) for each of the different charts with perturbations. For each type of chart, we showed two rows with charts. In the first row we showed copies of the original chart and in the second row we showed the charts with the added data perturbations which needed to be ranked from 1 to 9. This was done to have the same conditions when comparing the charts with perturbations with the respective original version. Each row of charts was randomly ordered so that there was no relation between the similarity and the position of the chart in the row. The data scientists completed the survey task at their discretion, but were asked to do so in maximum one day. Some data scientists returned the filled survey after a couple of hours and some on the next day. 

With the goal of analysing the visual similarity judgements and comparing them with the distance metrics for the different types of redundancy, we computed the average ranking per data perturbation for each of the 12 charts for all the data scientists.

\section{Comparison of the different redundancy types using different distance metrics}

In this section we compare the different types of redundancy for 9 distance metrics  using 3 approaches:
\begin{enumerate}
    \item Computing the $R^2$ coefficient of linear regressions per distance metric and redundancy type vs. the means of the Data Scientists rankings
  \item Computing the F1 score results for the top ranking (1) per distance metric and redundancy type vs. the rankings by the Data Scientists
    \item Computing the Mutual Information between the resulting rankings of each combination of distance metric and redundancy type and the average rankings by the Data Scientists  
    \end{enumerate}

We used the following distance metrics: Euclidean/Manhattan, Cosine, Jaccard, Dice, Mutual Information (with Mutual Information Diagrams), Jenssen-Shannon divergence and Pearson (Note: in the case of time series, the Euclidean distance is equivalent to the Manhattan distance, so we refer to both in all figures below as Man/Euc).

\subsection{Using  $R^2$ to compare different redundancy methods vis a vis the DS ranking}

For comparing the different combination of distance metrics and redundancy types, we used as benchmark the similarity rank for each modified chart we received from the data scientists survey.  Our goal, in this subsection, is to test how well the methods for detecting pattern similarity predict the data scientists' rankings. This can be done by running linear regressions between the ranking by the data scientists and the ranking resulting from each combination of distance metrics and redundancy types and selecting the best of the predictions (with the higher  $R^2$). 

For each of the 12 data sets (Original chart) and the 9 perturbations mentioned above we normalized the distances taken from the MIDs ($MID_nd$) and converted the similarity measures to distance measures ($distance = 1 - simmilarity$). 
After normalizing the distances for the different metrics, we transformed the scale, so that the minimum metric would be 1 and the maximum 9. We applied "PATRED" to each of the distance metrics with different numbers of redundancy points as mentioned above.

\begin{figure}[!ht]
\centering
\includegraphics[width =8.5 cm]{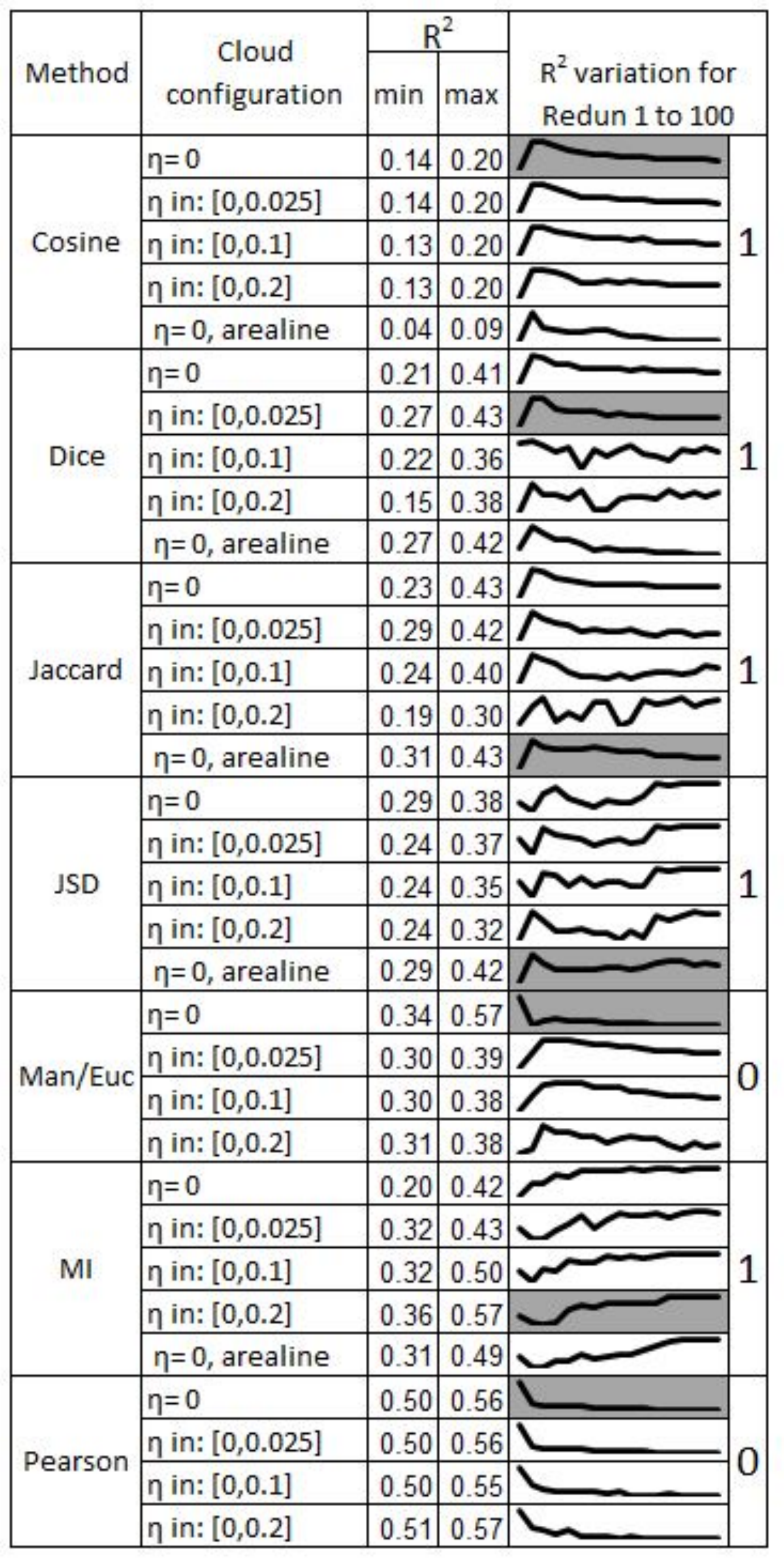}
\caption{Effects of the number of redundant points on $R^2$, the correlations of the rankings with the different distance metrics and the data scientists' rankings for the different redundancy types. Min and Max are the minimum and maximum values of $R^2$ within 0 to 100 redundant data points.  $\eta$ is the level of uniform random data points added to the line segments. Shadowed cells indicate best values.  0 indicates redundancy does not help, 1 indicates redundancy helps consistently.}
\label{R2}
\end{figure}

Figure ~\ref{R2} shows the effects of the redundancy configurations on the correlations with the data scientists' ratings $R^2$ for each one of the distance metrics. The sparklines on the right show the shape of the variation of $R^2$ for different numbers of redundancy points per line segments ([0,100]) together with cloud redundancy with different $\eta$ values.  The shadowed cells indicate the row with the best values. Not all the distance metrics were equally influenced by adding redundancy.  Below we summarize the effects of the different types of redundancy on the correlations. 

\begin{itemize}
    \item Redundancy did not help to improve $R^2$ for the Pearson and Man/Euc metrics. $R^2$ decreased after adding the first redundancy point.
     \item All other metrics improved their $R^2$ after adding 1 redundancy points, reaching the maximum at different numbers of redundancy points.
    \item For the Mutual Information and Jaccard metrics, the arealine redundancy had the best performance.
    \item For the JSD and Dice metrics, the arealine redundancy had slightly lower performance. 
    \item All metrics reached a steady state after a relatively low number of redundancy points.
    \item The Man/Euc and Pearson metrics are not appropriate for use with arealine redundancy as they can only work with vectors which have the same number of data points, and which are paired in a specific order (which is not the case with areas which use squared bins). 
\end{itemize}

\begin{figure}[ht]
  \centering
  \includegraphics[width = 8.5 cm]{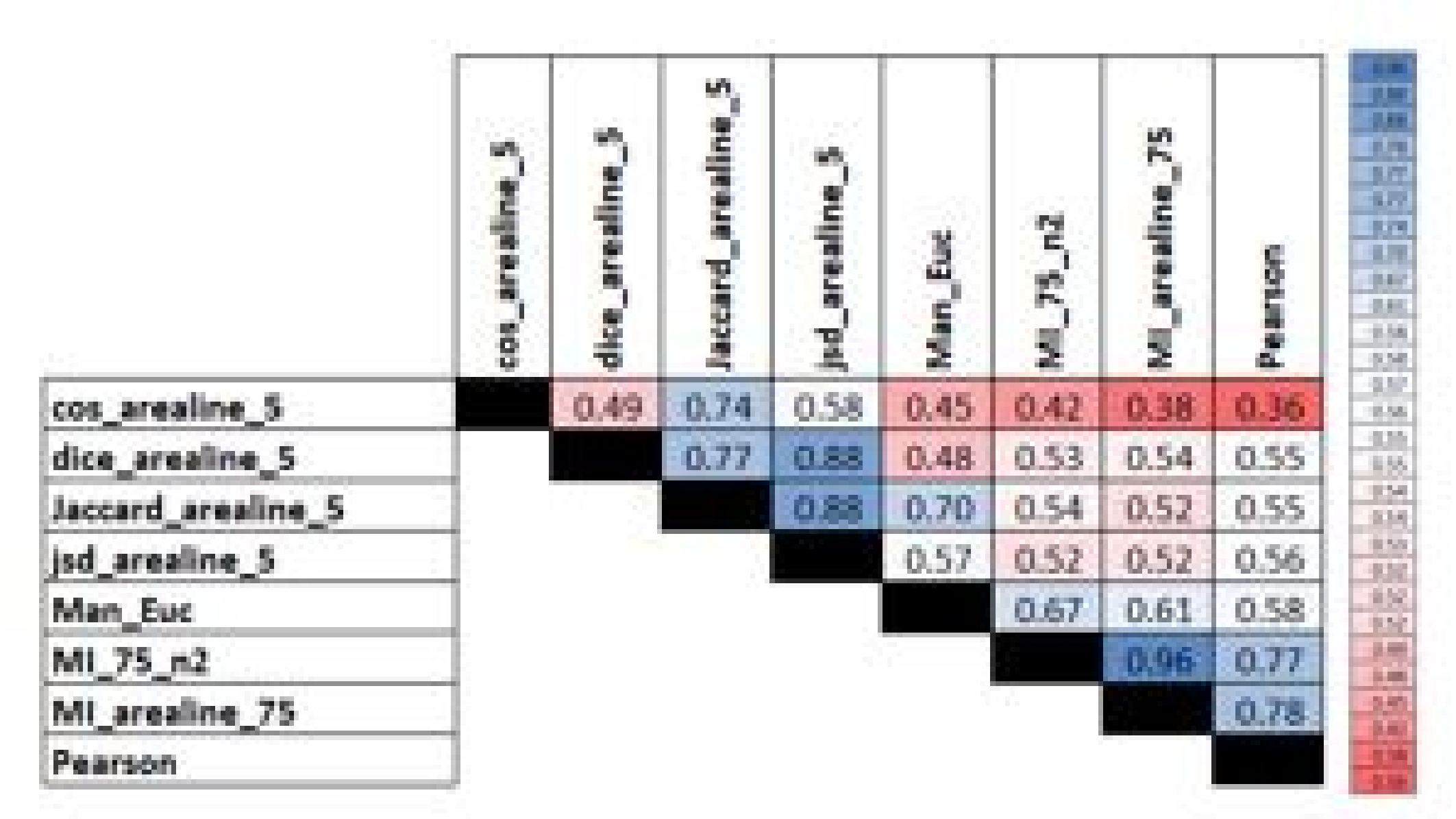}
  \caption{Correlations between the best redundancy configurations. The numbers after the underscore are the number of redundancy points and n2  refers to uniform noise with $\eta$ in [0,0.2]  }
 \label{correlations}
  \end{figure}

Figure ~\ref{correlations} shows the correlations between the rankings with the different best performing distance metrics. The blue cells group indicates the highly correlated distance metrics.

\subsection{Using the F1 score to compare different redundancy methods for the extreme cases}
In the previous analysis we compared the results for each of the ranks, without focusing on the extreme cases, when the charts are very similar or very different. Here we focus on the case when the rank is 1 and compare the different combinations of distance metrics and redundancy types from that perspective, using the F1 score, which is the harmonic mean of precision and recall. Precision answers the question of “what proportion of predicted positives are truly positive?”, and Recall answers the question of “what proportion of actual positives are correctly classified?”. It is easy to compute the two measures for binary classifications such as the top (1) rank. We did this by generating the confusion matrix for each of the cases, and we grouped the results into two groups: a) $rank=1$ and $rank\neq1$. Then we computed the harmonic mean of the precision and recall values to get the F1 scores. 

\begin{figure}[!ht]
\centering
\includegraphics[width =8.5 cm]{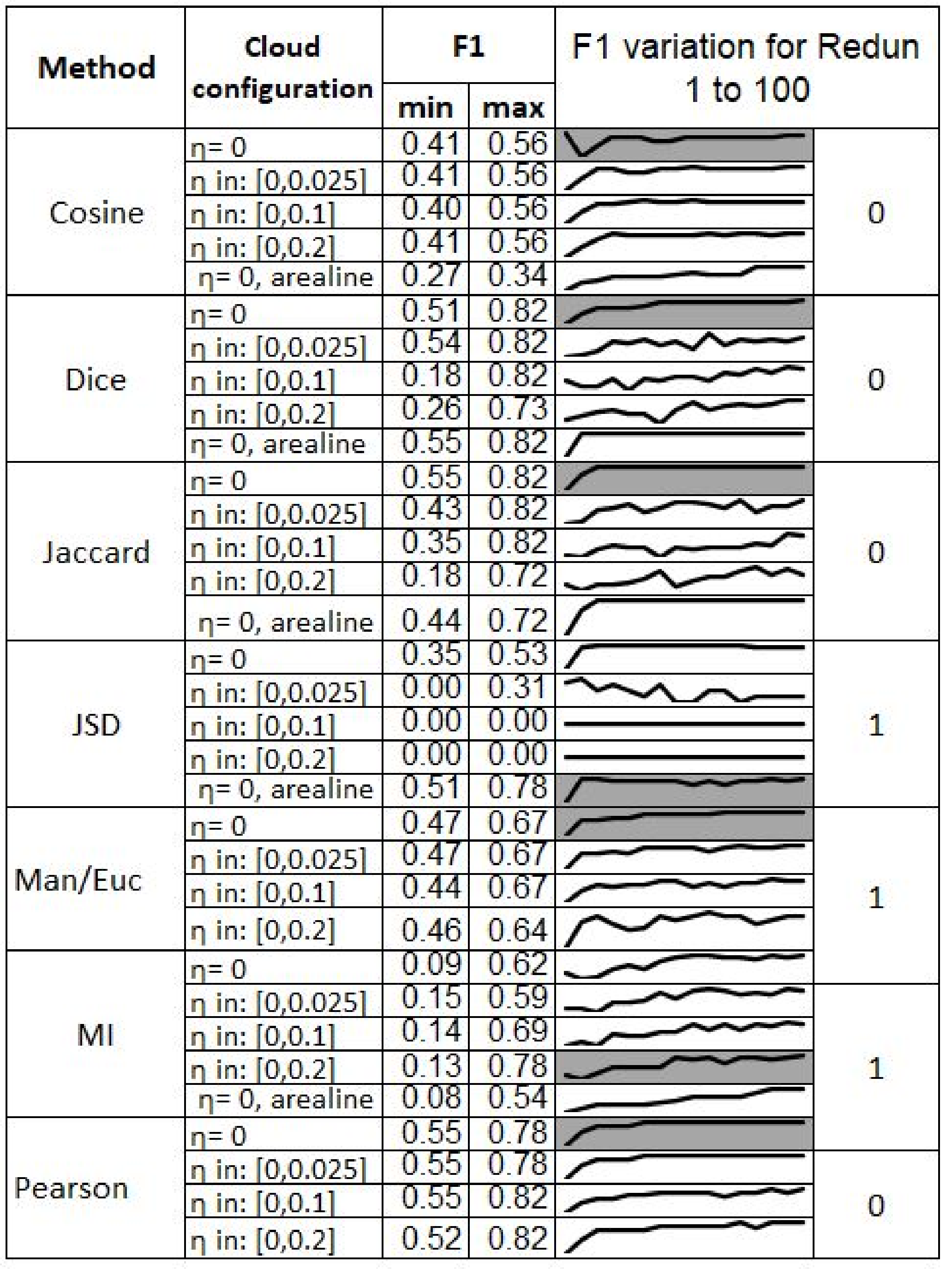}
\caption{Effects of the number of redundant points on F1 values for Rank 1. Min and Max are the minimum and maximum values of F1 within 0 to 100 redundant data points. The shadowed cells indicate the best values. $\eta$ is the level of uniform random data points added to the line segments.}
\label{meth_F1_1}
\end{figure}

Figure ~\ref{meth_F1_1} shows the effect of the redundancy types on the F1 score. Below we summarize the behavior of  F1 for the distance metrics with the different types of redundancy. For the Cosine, Dice, Jaccard and Pearson distance metrics, adding redundancy lowered the F1 score. Redundancy improved the F1 score for JSD (with area line), Manhattan/Euclidean (with equidistant points) and MI (with cloud).

\subsection{Using Normalized Mutual Entropy of sequences to compare the redundancy types with DS ranked charts}

\begin{figure}[!ht]
\centering
\includegraphics[width =8.5 cm]{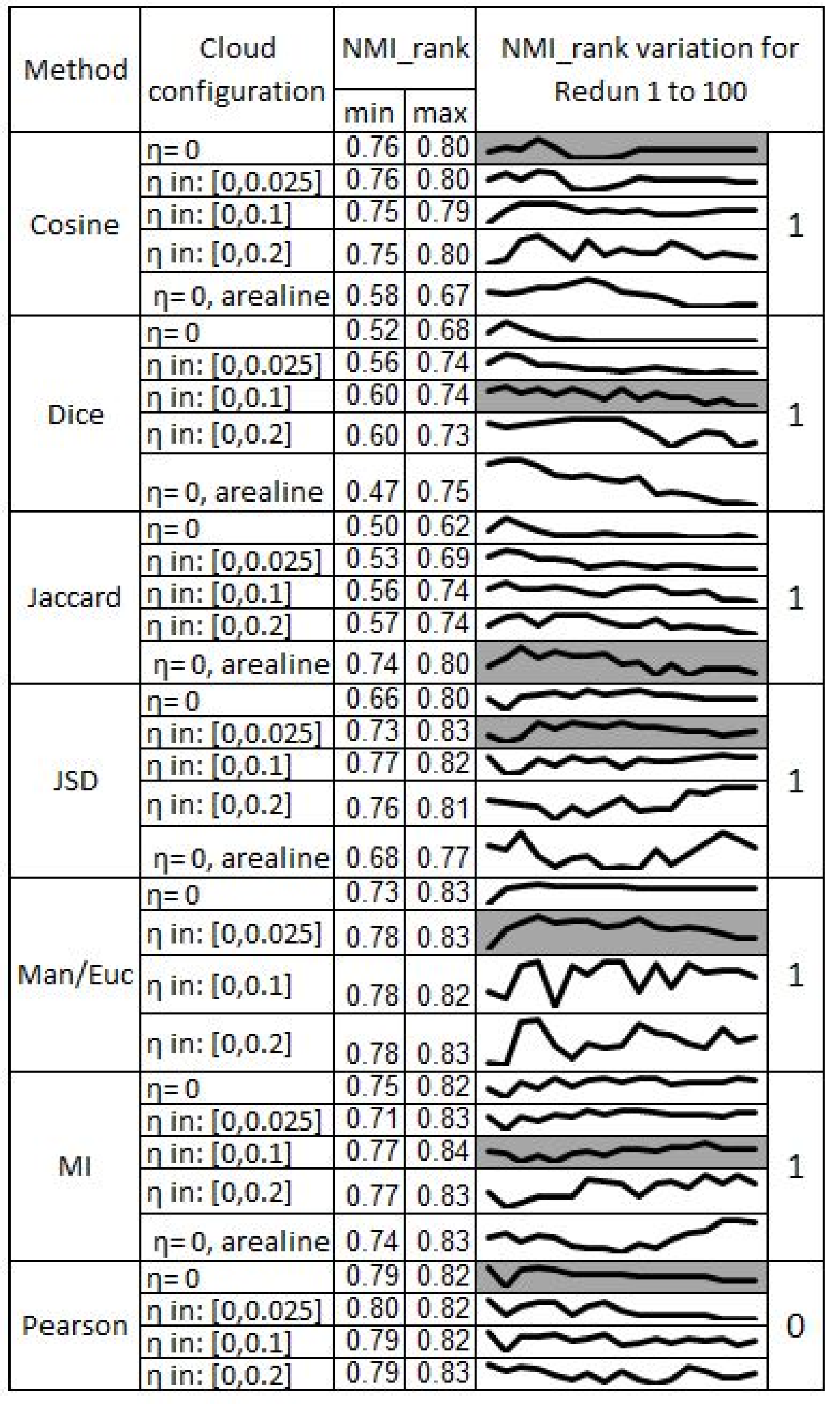}
\caption{Effects of the redundancy types on the similarity of the ranking produced by the different distance metrics and the ranking by the Data Scientists. Min and Max are the minimum and maximum values of the similarity within 0 to 100 redundant data points.  $\eta$ is the level of added uniform random cloud redundancy. Shadowed cells indicate best values, 0 - redundancy does not help, 1 redundancy can help. }
\label{ranking}
\end{figure}

The data scientists' rankings and the rankings computed with the different distance metrics are all sequences (from 1 to 9), so it is possible to compare the rankings by calculating the average Normalized Mutual Information of the sequences for all the charts. The method that most resembles the data scientists' rankings will have the highest Mutual Information.  We used the "DescTool" R package to compute the normalized Mutual Information for the different methods. As can be seen in Figure ~\ref{ranking}, redundancy helped to improve the Normalized Mutual Information for sequences for all methods, except Pearson.

\section{Summary of the comparison between the redundancy types, distance metrics and limitations}

\subsection{Summary}
We compared the similarity results for several distance metrics with and without redundancy (created with different redundancy types) to the "ground truth" that was based on data scientists' similarity judgements of visualizations of the same data sets. The Mutual Information, JSD, Jaccard, Dice and Cosine distance metrics had consistently higher $R^2$  and NMI values for sequences with redundancy than without it (sometimes the addition of only a single redundant data point raised the correlation). These distance metrics use data groupings to evaluate the similarity between data sets. The information theoretic methods (Mutual Information and JDS) use bins. Jaccard, Dice and Cosine use matrix calculations, applied to all matrix members concurrently. When we add redundant data points, they "integrate" with the original data, improving the Mutual Information to a certain point.

The Pearson, Euclidean and Manhattan methods use distances between corresponding pairs of individual data points or with the mean of the data-set. In this case, redundant additional data points can cause an increase in the sum of the differences (because we add more pairs of data points), possibly increasing the total distance between the data sets, which sometimes has a negative impact on the distance metrics.

Some distance metrics provide more or less similar results as can be seen with the cross-correlations between the different distance metrics rankings shown in Figure ~\ref{correlations}.  

Although redundancy had a lower effect on the F1 values for the top ranking (1), it still consistently helped the Shanon Entropy family methods (MI and JSD). 

\begin{figure*}[ht]
\centering
\includegraphics[width =18 cm]{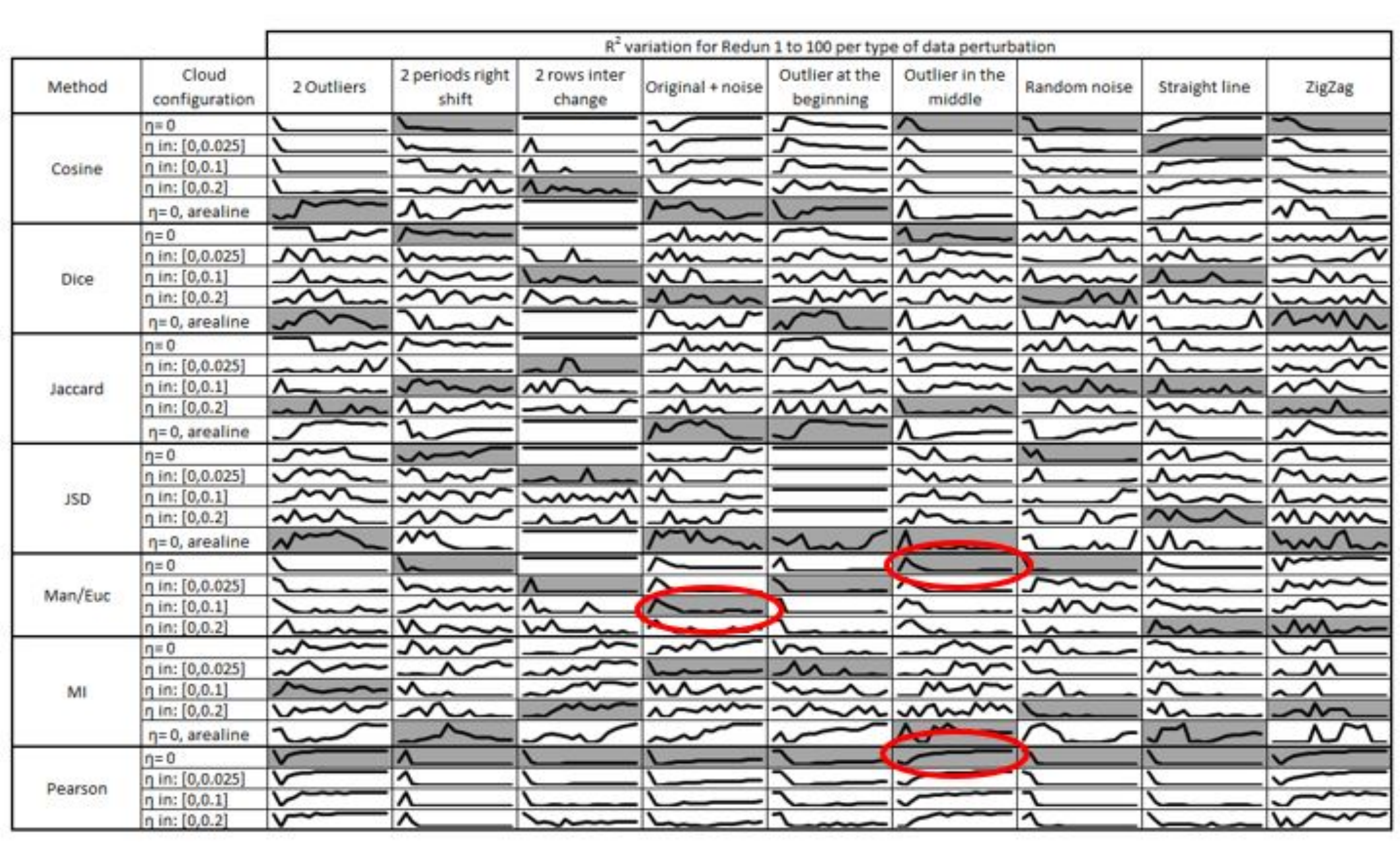}
\caption{Effects of redundancy types on $R^2$, per data perturbation. The red ellipses show when redundancy helped also for the Pearson and Manhattan/Euclidean distances. The shaded cells indicate the best performance (highest values of $R^2$)}.
\label{R2_type1}
\end{figure*}

The different data perturbation had different effects, as shown in Figure ~\ref{R2_type1}. Perturbations that change the position of the data points, such as shifting or interchanging, had the strongest negative impact on the different methods for measuring similarity. When the data perturbation involved only a single outlier in the middle of the pattern, most distance metrics improve already after adding one redundancy point. The exception was Pearson, which improves after adding more redundancy points. 

\subsection{Limitations}
The added redundancy approach can help in the search for visual patterns, but it has some limitations. First, we only tested specific line charts and with specific data perturbations, and a wider range of displays and data changes should be studied. Also, additional distance metrics should be studied. Finally, redundancy can be added in additional ways, and these, too should be studied.

\section{CONCLUSION AND DIRECTIONS FOR FUTURE RESEARCH}

\subsection{Conclusions}

Overall results indicate that adding visual redundancy to displayed data can help to detect similarities and differences between visualized patterns when using Intersection, Inner Product and Shannon's entropy family distances. Adding visual redundancy may not improve similarity detection when using distances from the Squared $L_2$ and $L_p$ Minkowski families, because these families of metrics are based on sums of distances between pairs of corresponding points in the two data sets. When redundant data is added, the sums of distances increase. This is not the case for the Intersection, Inner Product and Shannon's families where the metrics are mainly calculated at the whole data set level. However, for specific types of data perturbations (e.g. outliers positioned within the pattern) the Squared $L_2$ and $L_p$ Minkowski methods could also be enhanced with redundant data.

\begin{figure*}[!ht]
  \centering
  \includegraphics[width = 19cm]{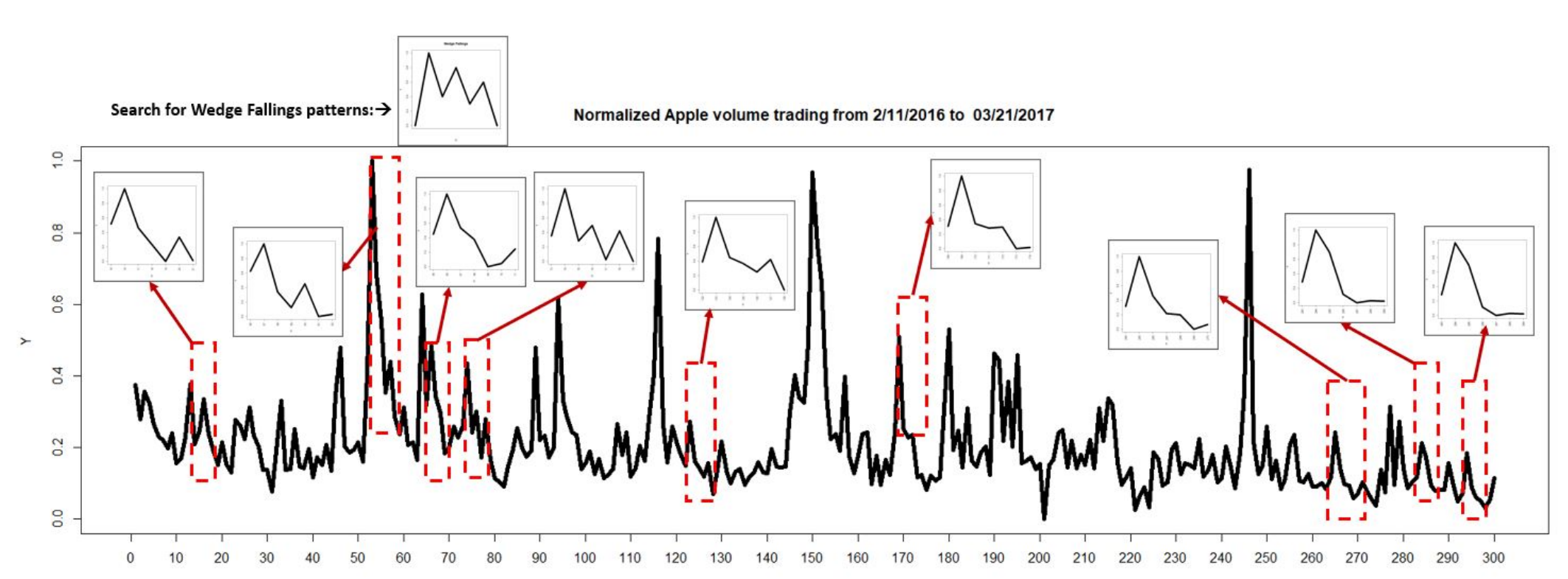}
  \caption{Search for Wedge Fallings pattern using redundant data}
 \label{how_to}
  \end{figure*}

\begin{figure}[ht]
  \centering
  \includegraphics[width = 9cm]{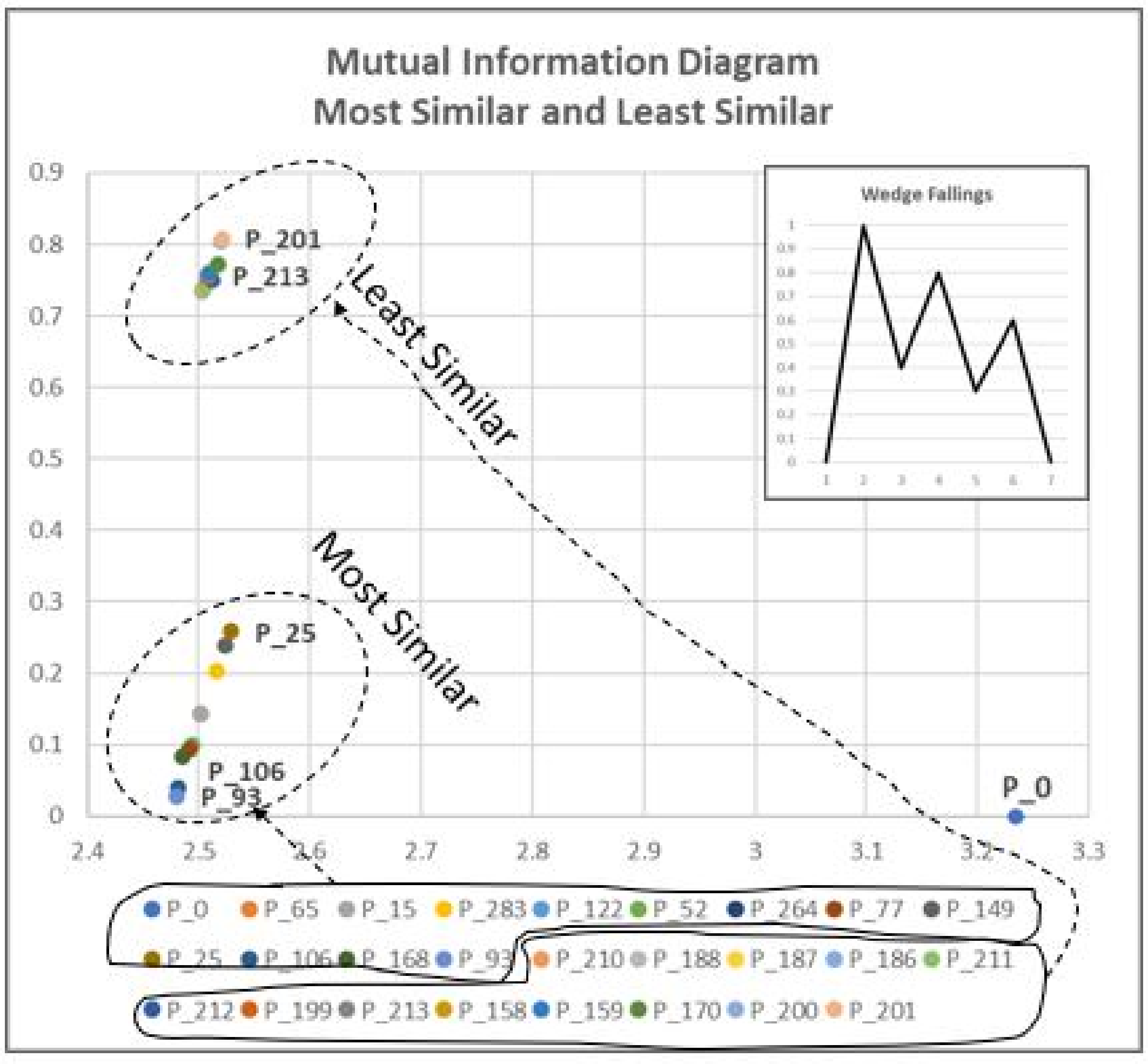}
  \caption{Mutual Information Diagram, showing the position of the most similar and least similar patterns to the Wedge Falling motif shown in Figure ~\ref{how_to}. P\_O is the position of the Wedge Falling pattern. The other points are for the identified patterns in the data, where the number is the starting point of the pattern in the data}
 \label{how_to1}
  \end{figure}

\subsection{Example of practical use}

Figure \ref{how_to} shows, for illustration, nine instances of similar motifs were identified when adding redundancy. The searched motif is of type Wedge Fallings ~\cite{wan2017formal}. In this case data we had the coordinates of the seven  points of the pattern from ~\cite{wan2017formal} and applied the 100 redundancy method for both the searched pattern and the data. If the sought pattern or motif is an expected pattern, then it should be digitized, either directly by providing the coordinates of each data point (as we did here) or by using a tablet, as depicted in Figure \ref{digitize1}. In our study, each time unit was of one day, and the total pattern is of 6 days (or 7 data points).  Each 7-days period in the data was normalized before performing the Mutual Information computations.  When looking for the same pattern for longer time units (e.g. weeks or months), one needs to smooth the data before doing the data redundancy and mutual information calculations. Figure \ref{how_to1} shows another way of looking at the identified patterns in the data, using a Mutual Information Diagram (MID), which can help to visually identify clusters.  Here we sorted the charts according to their Variation of Information (VI) and show their location in the MID for the top (most similar) and bottom (least similar) to the searched pattern.

\subsection{Future directions}

In this study we focused on pattern search and identification, using data redundancy that is consistent with the visual geometry of the data. The geometric consistency of the redundancy makes it possible to recover the original data fully, minimizing the added noise. One feasible future direction is to implement other geometries such as triangles.

This work can be expanded into several directions. The first is related to the identification of associations between variables, when the relation between them is not linear. A second direction for future research is anomaly  detection in multi-variate variable time series by trying to identify what is causing strong decreases of Mutual Information when moving along the time axis or while moving along a radial axis for 2D images. 

To summarize, the adding of "redundant visual information" can provide significant benefits when analysing patterns in the data, and it opens some promising new directions for further research.

\section{Appendix}

The simplest two distance metrics between variables are the Manhattan and Euclidean distances ~\cite{ljubesic2008comparing}.
Manhattan distance (Equation ~\ref{eqmnhtn}) calculates the distance between vectors on all dimensions, while Euclidean distance (Equation ~\ref{eqeuc}) measures the geometric distance between the two vectors. The Euclidean distance is sensitive to extreme values, i.e., outliers, in the variables ~\cite{kaufman2009finding,lee2000measures,ljubesic2008comparing}.

\begin{equation}\label{eqmnhtn}
\ Manhattan\; Dist_{Mnhtn}(X,Y) = \sum_{i=1}^{N}|X_{i}-Y_{i}|
\end{equation}

\begin{equation}\label{eqeuc}
\ Euclidean\; Dist_{Euc}(X,Y) = \sqrt{\sum_{i=1}^{N}(X_{i}-Y_{i})^2}
\end{equation}

The Cosine distance (Equation ~\ref{eqcos}) uses the dot product operator from linear algebra ~\cite{ljubesic2008comparing} and is equal to the cosine between two vectors  ~\cite{ljubesic2008comparing}. It is a common measure in information retrieval research ~\cite{rahutomo2012semantic}. 

\begin{equation}\label{eqcos}
\ Cosine\; d_{Cos}(X,Y) =  1 - \frac{X \cdot Y}{{|X| * |Y|}} 
\end{equation}

The Jaccard distance (Equation ~\ref{eqjacc}) is broadly used in computing string similarities ~\cite{ljubesic2008comparing}. The terms within $| |$ mean the number of elements that are shared and the total number of distinct elements. The measure compares members for two sets to see which members are shared and which are distinct. It is a measure of the similarity of two sets of data, with a range between 0 and 1. For two identical sets, the distance will be 0; when they are completely different (no common elements) the distance will be 1.  The measure is easy to interpret, but it may be sensitive to small sample sizes and may give erroneous results ~\cite{123StephanieGlen}. 

\begin{equation}\label{eqjacc}
\ Jaccard \ distance \; d_{JC}(X,Y) = 1 - \frac{|X \cap Y|}{|X \cup Y| }
\end{equation}

The Jaccard distance has an intuitive interpretation, which links it to Mutual Information.  Its similarity complement (1 - $d_{JC}$) is related to the measure of the probability that an element of at least one of two sets is an element of both, and thus is a reasonable measure of similarity or the intersection between the two ~\cite{levandowsky1971distance}. 
The Dice distance (Equation ~\ref{eqdice}) is similar to the Jaccard measure and is also used in information retrieval ~\cite{ljubesic2008comparing}.  

\begin{equation}\label{eqdice}
\ Dice \ distance\; d_{Dice}(X,Y) = 1- \frac{2*|X \cap Y|}{|X| + |Y| }
\end{equation}

The Pearson correlation coefficient  ($r$, in the Squared $L_2$ family, Equation ~\ref{eqpear}) is often used to quantify the strength of the association between 2 variables, and its sign tells the direction of this association  ~\cite{glantz2001primer}. 

\begin{equation}\label{eqpear}
\ Pearson\; S_{r}(X,Y) =  \frac{\sum_{i=1}^{N}(\overline{X} - X_{i})(\overline{Y}-Y_{i})}{\sqrt{\sum_{i=1}^{N}(X_{i} - \overline{X} ) ^2} \sqrt{\sum_{i=1}^{N}(Y_{i} - \overline{Y})^2}}
\end{equation}

Since the Pearson coefficient $r$ is a similarity measure with values between -1 and 1, it is necessary to transform it into a normalized distance as in Equation ~\ref{dpear}.

\begin{equation}\label{dpear}
\ Pearson\ distance\; {d_r = (1 - r)/2}
\end{equation}

Pearson correlation and distance are very intuitive, and they can easily be used to evaluate linear associations. It is best for continuous, normally distributed data, but it is strongly affected by extreme values \cite{clark2013comparison}.

\ifCLASSOPTIONcompsoc

\fi

\ifCLASSOPTIONcaptionsoff
 \newpage
\fi

\bibliographystyle{abbrv-doi}
\bibliography{template}
% if you will not have a photo at all:

\begin{IEEEbiography} [{\includegraphics[width=1in,height=1.4in,clip,keepaspectratio]{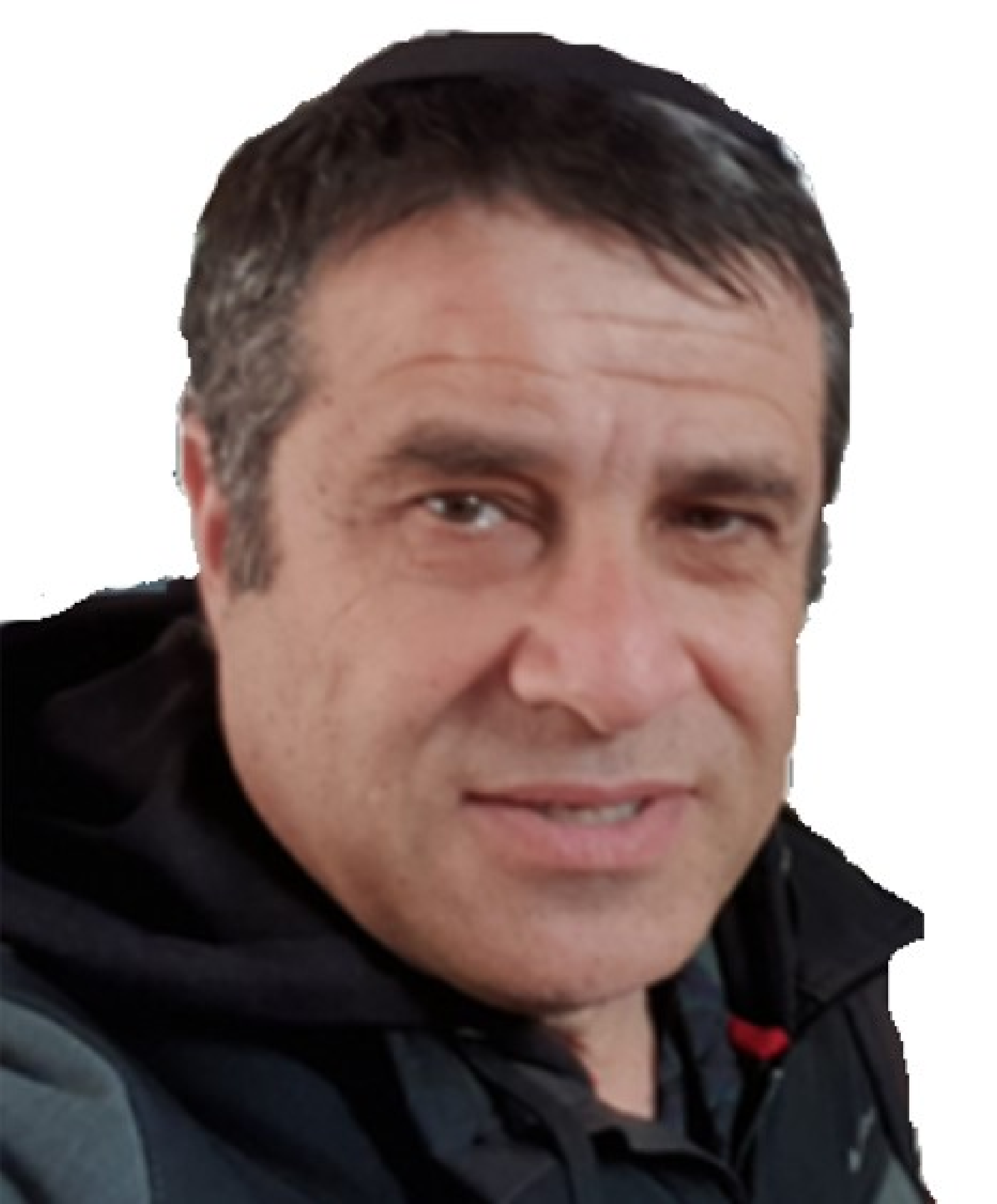}}]{Salomon Eisler}
is a Ph.D. student at the Department of Industrial Engineering at Tel Aviv University, Israel. He holds a M. Sc. in Electrical Engineering from the Technion - Israel Institute of Technology. His research interests are in Visualizations, Information Theory and Individual Differences. 
\end{IEEEbiography}

\begin{IEEEbiography}
[{\includegraphics[width=1in,height=1.25in,clip,keepaspectratio]{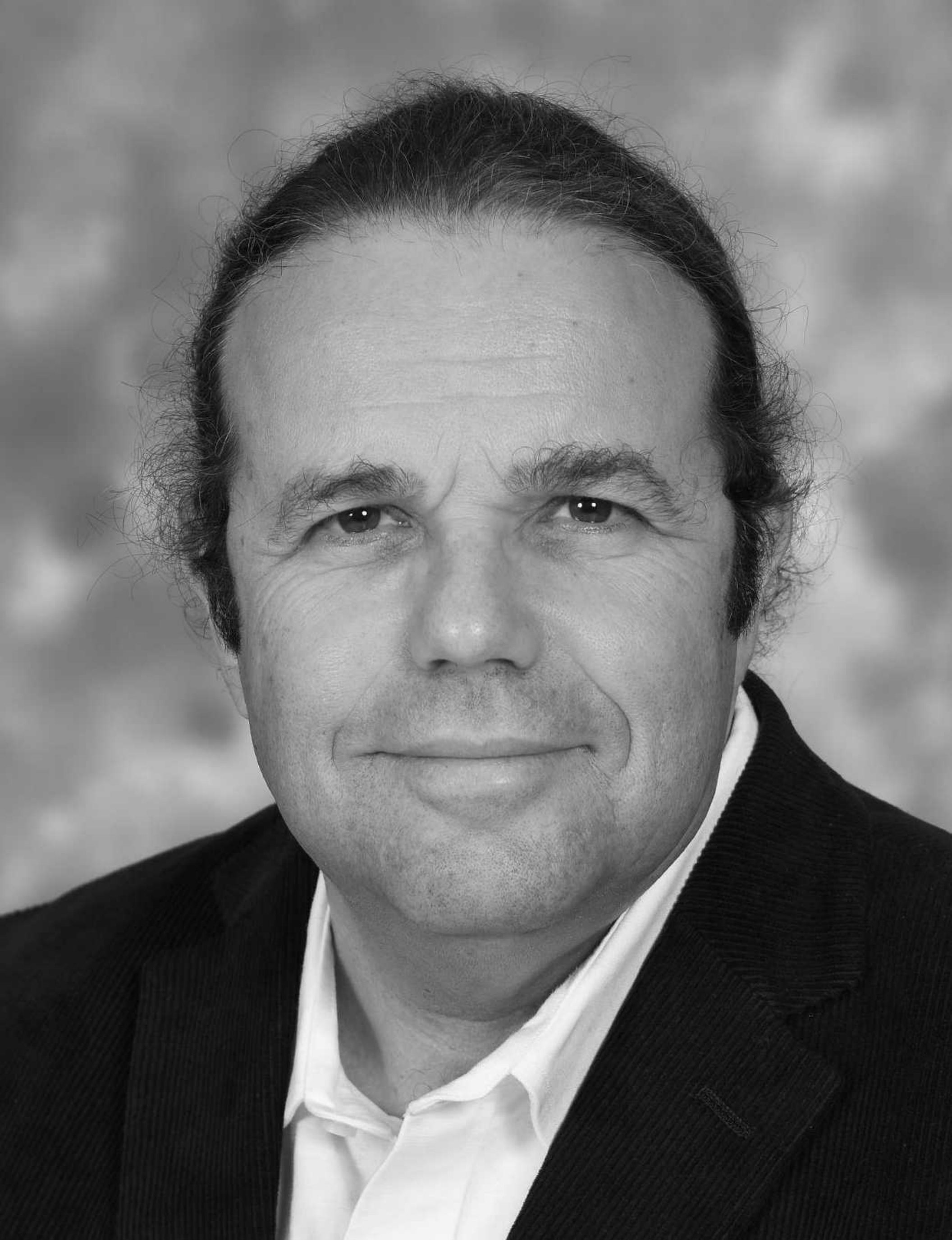}}]{Joachim Meyer}
 (M’01-SM’10) received an M.A. in psychology (in 1986) and a Ph.D. in Industrial Engineering (in 1994) from Ben-Gurion University of the Negev, Israel. He held research positions at the Technion – Israel Institute of Technology, Harvard University and the Massachusetts Institute of Technology, was on the faculty of Ben-Gurion University of the Negev, and is currently the Celia and Marcos Maus Professor for Data Sciences at the Department of Industrial Engineering at Tel Aviv University, Tel Aviv, Israel. He is a fellow of the Human Factors and Ergonomics Society and served for many years as an Associate Editor or member of the editorial board for IEEE Transactions and other journals.
\end{IEEEbiography}

\end{document}